\documentclass[acmsmall]{acmart}
\usepackage{bbm}
\usepackage[normalem]{ulem}
\usepackage{bm}
\usepackage{amsmath}
\usepackage[french,vlined,linesnumbered,ruled]{algorithm2e}
\usepackage{etoolbox}
\usepackage{enumitem}
\usepackage{graphicx}
\usepackage{footmisc}
\usepackage{subcaption}

\newtheorem{definition}{DEFINITION}
\newtheorem{theorem}{THEOREM}
\newtheorem{Corollary}{COROLLARY}

\makeatletter
\let\std@footnotetext\@footnotetext
\usepackage{setspace}
\let\@footnotetext\std@footnotetext
\makeatother
\AtBeginDocument{%
  }

\setcopyright{acmlicensed}
\acmJournal{PACMMOD}
\acmYear{2025} \acmVolume{3} \acmNumber{6 (SIGMOD)}
\acmArticle{318} \acmMonth{12}
\acmDOI{10.1145/3769783}

\received{April 2025}
\received[revised]{July 2025}
\received[accepted]{August 2025}
\begin{document}

\title{Dynamically Detect and Fix Hardness for Efficient Approximate Nearest Neighbor Search}

\author{Zhiyuan Hua}
\email{huazy@nbjl.nankai.edu.cn}
\orcid{0009-0006-3222-0544}
\affiliation{%
  \institution{College of Computer Science, Nankai University}
  \city{Tianjin}
  \country{China}
}

\author{Qiji Mo}
\email{moqj@nbjl.nankai.edu.cn}
\orcid{0009-0003-6931-6396}
\affiliation{%
  \institution{College of Computer Science, Nankai University}
  \city{Tianjin}
  \country{China}
}

\author{Zebin Yao}
\email{yaozb@nbjl.nankai.edu.cn}
\orcid{0000-0001-7418-5691}
\affiliation{%
  \institution{College of Computer Science, Nankai University}
  \city{Tianjin}
  \country{China}
}

\author{Lixiao Cui}
\email{cuilx@nbjl.nankai.edu.cn}
\orcid{0000-0002-4017-0974}
\authornote{corresponding authors}
\affiliation{%
  \institution{College of Computer Science, Nankai University}
  \city{Tianjin}
  \country{China}
}

\author{Xiaoguang Liu}
\email{liuxg@nbjl.nankai.edu.cn}
\orcid{0000-0002-9010-3278}
\authornotemark[1]
\affiliation{%
  \institution{College of Computer Science, Nankai University}
  \city{Tianjin}
  \country{China}
}

\author{Gang Wang}
\email{wgzwp@nbjl.nankai.edu.cn}
\orcid{0000-0003-0387-2501}
\affiliation{%
  \institution{College of Computer Science, Nankai University}
  \city{Tianjin}
  \country{China}
}

\author{Zijing Wei}
\email{wzjrj28@outlook.com}
\orcid{0009-0005-8488-9314}
\affiliation{%
  \institution{Alibaba Group Holding Limited}
  \city{Beijing}
  \country{China}}

\author{Xinyu Liu}
\email{lxy264173@alibaba-inc.com}
\orcid{0000-0002-0292-1351}
\affiliation{%
  \institution{Alibaba Group Holding Limited}
  \city{Beijing}
  \country{China}}

\author{Tianxiao Tang}
\email{tianxiao.ttx@alibaba-inc.com}
\orcid{0009-0007-2585-8803}
\affiliation{%
  \institution{Alibaba Group Holding Limited}
  \city{Beijing}
  \country{China}}

\author{Shaozhi Liu}
\email{shaozhi.liu@gmail.com}
\orcid{0009-0008-1711-1328}
\affiliation{%
  \institution{Alibaba Group Holding Limited}
  \city{Beijing}
  \country{China}}

\author{Lin Qu}
\email{xide.ql@taobao.com}
\orcid{0009-0004-2028-0780}
\affiliation{%
  \institution{Alibaba Group Holding Limited}
  \city{Beijing}
  \country{China}}

\renewcommand{\shortauthors}{Zhiyuan Hua et al.}

\begin{abstract}
Approximate Nearest Neighbor Search (ANNS) has become a fundamental component in many real-world applications. Among various ANNS algorithms, graph-based methods are state-of-the-art. However, ANNS often suffers from a significant drop in accuracy for certain queries, especially in Out-of-Distribution (OOD) scenarios. To address this issue, a recent approach named RoarGraph constructs a bipartite graph between the base data and historical queries to bridge the gap between two different distributions. However, it suffers from some limitations: (1) Building a bipartite graph between two distributions lacks theoretical support, resulting in the query distribution not being effectively utilized by the graph index. (2) Requires a sufficient number of historical queries before graph construction and suffers from high construction times. (3) When the query workload changes, it requires reconstruction to maintain high search accuracy.

In this paper, we first propose Escape Hardness, a metric to evaluate the quality of the graph structure around the query. Then we divide the graph search into two stages and dynamically identify and fix defective graph regions in each stage based on Escape Hardness. (1) \textbf{From the entry point to the vicinity of the query.} We propose \textbf{R}eachability \textbf{Fix}ing (RFix), which enhances the navigability of some key nodes. (2) \textbf{Searching within the vicinity of the query}. We propose \textbf{N}eighboring \textbf{G}raph Defects \textbf{Fix}ing (NGFix) to improve graph connectivity in regions where queries are densely distributed. The results of extensive experiments show that our method outperforms other state-of-the-art methods on real-world datasets, achieving up to 2.25× faster search speed for OOD queries at 99\% recall compared with RoarGraph and 6.88× faster speed compared with HNSW. It also accelerates index construction by 2.35-9.02× compared to RoarGraph.
\end{abstract}

\begin{CCSXML}
<ccs2012>
   <concept>
       <concept_id>10002951.10003317.10003325</concept_id>
       <concept_desc>Information systems~Information retrieval query processing</concept_desc>
       <concept_significance>500</concept_significance>
       </concept>
   <concept>
       <concept_id>10002951.10002952.10003190.10003192.10003210</concept_id>
       <concept_desc>Information systems~Query optimization</concept_desc>
       <concept_significance>500</concept_significance>
       </concept>
 </ccs2012>
\end{CCSXML}

\ccsdesc[500]{Information systems~Information retrieval query processing}
\ccsdesc[500]{Information systems~Query optimization}

\keywords{Graph-based Index, High-Dimensional Vector, Approximate Nearest Neighbor Search}

\maketitle

\section{INTRODUCTION}
Nearest Neighbor Search (NNS) plays a crucial role in many real-world applications, such as large-scale information retrieval ~\cite{webvid,glove,ir3}, recommendation systems~\cite{rs,rs2}, and Retrieval Augmented Generation (RAG) in large language models (LLM)~\cite{rag1,rag2,rag3}. These applications require fast response times, but exact NNS in high-dimensional spaces is unable to meet the practical demands of current applications. As a result, most applications now rely on Approximate Nearest Neighbor Search (ANNS) for efficient neighbor search. To ensure high accuracy while maintaining low latency in ANNS, numerous methods have been proposed~\cite{tree1,tree3,pq,pq_survey,rabitq,pm-lsh,db-lsh,det-lsh,ad_sampling,lsh_apg,hnsw,roar_graph,nsg,t_mg,symphonyQG,rabitq2,tribase,graph_survey,subspace,finger,learing_to_route,grasp,Reinforcement_route,elpis,starling}. Among these, graph-based methods currently achieve the best performance in many scenarios~\cite{survey, hnsw, nsg}.

However, graph-based methods often suffer from a significant drop in accuracy for certain queries~\cite{early_stop, steiner_hardness} (referred to as 'hard queries' in this paper). We also observed this phenomenon in the real production environment of our e-commerce platform, which significantly affects the user experience. The main reasons for the accuracy degradation are as follows: (1) Graph-based indexes are typically approximations of Relative Neighborhood Graph (RNG)~\cite{rng} or its variants ~\cite{nsg,ssg,t_mg}, and theoretical guarantees on search accuracy only hold when the queries overlap with or are highly similar to the base data. However, in practice, not all queries are close to the base data, and queries that are farther from the base data tend to have lower accuracy. (2) Graph-based methods typically utilize a greedy search algorithm that explores only the closest vectors to the query at each step. Without effective edges in the graph index, the search may only reach a subset of the Nearest Neighbors (NNs), leading to a drop in accuracy. In particular, the drop in accuracy is more frequent in cross-modal retrieval scenarios due to the existence of the modality gap~\cite{ModalityGap}. This gap causes the distributions of query vectors (e.g., texts) and base vectors (e.g., images) to differ significantly (i.e., query vectors are Out-of-Distribution (OOD)), making queries distant from the base data~\cite{roar_graph}. (The formal definition of OOD queries will be provided in Section \ref{preliminaries}.)

To improve index performance on hard queries, especially in OOD scenarios, recent methods~\cite{roar_graph,ood_diskann} aim to bridge the distribution gap between base data and queries. Among them, RoarGraph~\cite{roar_graph} is a representative method. The index construction method of RoarGraph consists of the following steps: (1) RoarGraph builds a bipartite graph between the base data $X$ and the historical query set $Q$ to connect the two distributions. (2) It then projects the query points onto the base data (using the nearest neighbor of query to replace the query points) to avoid adding query points directly into the graph. (3) Finally, it enhances graph connectivity by assigning more neighbors to each node. RoarGraph constructs a high-quality graph index through the above approach, which helps greedy search discover more NNs.

\begin{figure}
    \centering
    \setlength{\abovecaptionskip}{0.1cm}
    \includegraphics[width=0.7\linewidth]{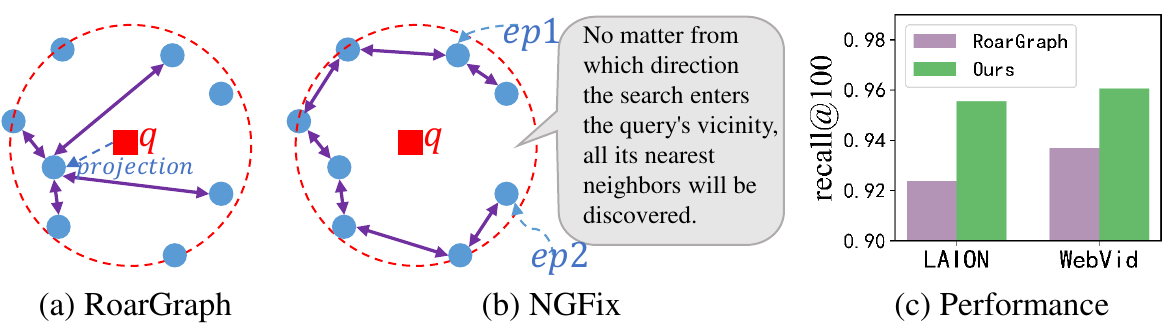}
    \caption{Comparison between RoarGraph and Our Methods. (a) Example of RoarGraph. (b) Example of NGFix. (c) Query accuracy under same average latency in LAION10M\cite{laion} and WebVid2.5M\cite{webvid}.}
    \Description{..}
    \label{query_example}
    % \vspace{-1em}
\end{figure}

Despite the significant performance improvements achieved by RoarGraph, it still has the following limitations: (1) Building a bipartite graph between two distributions lacks theoretical support. As a result, some effective edges are overlooked by RoarGraph. As shown in Figure \ref{query_example}a, RoarGraph fails to fully utilize the information provided by historical queries, resulting in some nodes not being connected to their most effective neighbors when compared to Figure \ref{query_example}b. (2) It often relies on a substantial number of historical queries during the graph construction phase to achieve satisfactory performance. It also requires calculating the exact $k$NN for each historical query during graph construction, making it suffer from long construction times. (3) When the query workload changes, RoarGraph requires a full reconstruction to maintain accuracy. However, in real-world production environments, the hard queries are constantly changing, making it difficult for RoarGraph to adapt to such changes. For example, on our e-commerce platform, a comparison between query workloads from two different time periods shows that approximately 10\% of the queries in the newer period are far from the distribution of the previous queries.

To address the limitations above, the key idea of our methods is to leverage online queries to dynamically fix defects of the graph in the regions where the queries are densely distributed. Due to the varying hardness of queries on a given graph index~\cite{early_stop,steiner_hardness}, we propose \textbf{Escape Hardness} (EH), a theoretically guaranteed method for quantifying query hardness. Following previous work ~\cite{steiner_hardness, scc}, we divide the search into two phases: (1) From the entry point to the vicinity of the query. (2) Searching within the vicinity of the query. For most queries, greedy search easily enters the second phase ~\cite{steiner_hardness} (solution of some counterexamples will be introduced later) but often reaches only a part of the NNs. Therefore, we primarily measure the quality of the graph structure around each query. To avoid the impact of the entry point, we use a matrix to represent the difficulty of the query, where each element of the matrix indicates the difficulty of exploring from one point around the query to another point. Meanwhile, for a given query, if the values in the matrix are all small, the search accuracy will be guaranteed.

Based on EH, we propose \textbf{R}eachability \textbf{Fix}ing (RFix) and \textbf{N}eighboring \textbf{G}raph Defects \textbf{Fix}ing (NGFix) to dynamically fix defects of the graph. (1) RFix mainly focuses on the first phase of the search. For a historical query, if the search does not reach the query's vicinity, RFix first obtains the approximate nearest neighbor found by the search, and then enhances the navigation performance of the approximate nearest neighbor point by expanding its candidate neighbor set. (2) NGFix mainly focuses on the quality of the graph index in the second phase of the search. Guided by EH, it adds effective edges around the query to ensure that for each historical query, the difficulty of reaching other points from any point around the query is low. As shown in Figure \ref{query_example}b, NGFix ensures that greedy search finds all NNs of the query from any direction.  After applying NGFix and RFix, we guarantee the search accuracy for historical queries. The results of experiments shown in Figure \ref{query_example}c demonstrate that our method also performs well on test queries (different from historical queries). Meanwhile, to reduce the overhead of index construction, we replace exact $k$NN with approximate $k$NN during NGFix and RFix. Our contributions are summarized as follows:

\begin{itemize}[leftmargin=0.4cm, itemindent=0cm]
\item[$\bullet$] We first provide some theoretical analysis of the relationship between query accuracy and the local graph structure around the query. We then propose Escape Hardness to measure the difficulty of accessing other points from a given point near the query, thereby measuring the quality of the graph structure around the query. The results of experiments show that Escape Hardness is highly correlated with the actual query accuracy. 
\item[$\bullet$] We propose NGFix and RFix, using Escape Hardness in combination with the query distribution to dynamically detect and fix the defective areas of the graph. 
\item [$\bullet$] We conduct comprehensive experiments on real-world datasets, and the results demonstrate that our method significantly outperforms existing methods in both OOD and In-Distribution (ID) workloads, achieving up to 2.25× faster search speed for OOD queries at a 99\% recall rate and accelerating index construction by 2.35-9.02×. Meanwhile, compared to RoarGraph, we achieve the same search efficiency using only 8\%-30\% of the historical queries it requires.
\item [$\bullet$] We release a new cross-modal dataset to facilitate research in ANNS, which is collected from the real-world production environment of a large-scale e-commerce platform. We also provide a well-optimized C++ library for ANN query based on our method \footnote{\href{https://github.com/yuhuifishash/NGFix}{https://github.com/yuhuifishash/NGFix}}.
\end{itemize}

\section{PRELIMINARIES}
\label{preliminaries}

\textbf{ANNS.} In this paper, we let $X = \{x_1,x_2,...,x_n\} \subseteq \mathbb{R}^d$ be a finite set of $n$ vectors in a $d$-dimensional space. Let $\delta(\cdot,\cdot)$ denote the distance between two points and $N_{i,q}$ denote the $i$-th NN of $q$. For a given query $q \in \mathbb{R}^d$, the objective of the Approximate $k$ Nearest Neighbor Search (A$k$NNS) algorithm is to find a point set $S$, where $S$ satisfies $|S|=k$, $\delta(x, q) \le (1+\epsilon)\delta(x',q)$ for $x \in S$, $x' \in X\setminus S$, and $\epsilon$ is a small constant satisfied $\epsilon \ge 0$. We usually use recall and relative distance error (rderr) to evaluate the accuracy of query $q$. For a query $q$, $recall@k = {|S \cap KNN(q)|}/{k}$, where $KNN(q)$ represents the exact $k$NN of the query $q$; $rderr@k=avg_{i=1}^k({\delta(ANN_{i,q},q)}/{\delta(N_{i,q},q)}-1)$, where $ANN_{i,q}$ denotes the $i$-th approximate NN of $q$.

\begin{algorithm}
\small
\setstretch{0.8} 
\caption{Greedy Search}\label{beamsearch}
\KwIn{Graph index $G$, query $q$, result size $k$, entry point $ep$, search list size $L \ge k$.}
\KwOut{$k$NN of $q$}
Initialize a candidate set $\bm{C} \leftarrow \{ep\}$  and a result set $\bm{R} \leftarrow \{ep\}$\\
mark $ep$ as visited\\
\While{$\bm{C} \ne \emptyset$}{
    $u \leftarrow $ pop an element from $\bm{C}$ with minimum $\delta(u,q)$\\
    $d_{max} \leftarrow max_{x \in \bm{R}}\delta(x,q)$ \tcp*[f]{If $\bm{R} = \emptyset$, $d_{max} \leftarrow inf$}\\
    \textbf{if} $\delta(u,q) > d_{max} $ \textbf{then} break \\
    \For{$v \in Neighbors(G,u)$ and $v$ has not been visited}{
        mark $v$ as visited\\
        \If{$\delta(v,q) < d_{max}$ or $|\bm{R}| < L$}{
            $\bm{R} \leftarrow \bm{R} \cup \{v\}$,  $\bm{C} \leftarrow \bm{C} \cup \{v\}$ 
        }
        \While{$|\bm{R}| > L$}{
            pop an element $x$ from $\bm{R}$ with maximum $\delta(x,q)$
        }
        $d_{max} \leftarrow max_{x \in \bm{R}}\delta(x,q)$
    }
}
\Return closest $k$ points in $\bm{R}$

\end{algorithm}

\textbf{Graph Index.} We use a directed graph $G=(V,E)$ to represent a graph index, where $V$ is the set of vertices, and each vector in $X$ corresponds to a vertex in $V$. The set $E$ represents the edges, where each edge $(u,v)\in E$ indicates a proximity relationship between vertices $u$ and $v$. We let $Neighbors(G,v)$ denote the set of neighbors of vertex $v$ in graph $G$.

\textbf{Greedy Search.} We typically use Greedy Search for ANNS on graph indexes.  The details of the algorithm are presented in Algorithm \ref{beamsearch}. We begin by initializing a candidate set $\bm{C}$ and a result set $\bm{R}$, where $\bm{R}$ has a maximum length of $L$ (line 1). During the search, at each step, we select the point from $\bm{C}$ that is closest to the query $q$ and check its neighbors. If any neighbor is closer to the query than a point in $\bm{R}$, we add it to both $\bm{R}$ and $\bm{C}$ (lines 7-13). The search terminates when the distance from the current point to $q$ is greater than that of any point in $\bm{R}$ or when the $\bm{C}$ becomes empty (lines 3-6). Finally, the algorithm returns the $k$ closest points to the $q$ from $\bm{R}$. Since this method is an approximate algorithm, Greedy Search often terminates after exploring only a portion of the query’s nearest neighbors for certain queries, leading to reduced search accuracy. A common and simple approach to address this is to increase the search list size $L$, but this will lead to an increase in search latency. A more effective solution is to design an improved graph index that guides the greedy search more effectively, which is the main focus of this paper.

\textbf{OOD Queries.} In cross-modal ANNS, the base and query vectors are typically generated by a multimodal model (e.g., CLIP~\cite{clip}). Taking image-text retrieval as an example, image and text embeddings are generated by different encoders, and then these embeddings are mapped into a shared space through contrastive learning~\cite{clip}. However, there exists a modality gap between image and text vectors, so their distributions are different. A query is considered OOD if its modality differs from that of the base data. We can use the Wasserstein distance~\cite{wasserstein_distance} to mathematically measure the difference between the two distributions and use the Mahalanobis distance~\cite{Mahalanobis} to measure the distance from a vector to a distribution~\cite{roar_graph}.

\section{RELATED WORK}
The algorithm of ANNS can be broadly categorized into tree-based~\cite{tree1,tree2,tree3}, hashing-based~\cite{lsh,db-lsh,det-lsh,pm-lsh}, quantization-based~\cite{pq,pq_survey,rabitq,rabitq2,opq,scann,fast_scan}, and graph-based methods~\cite{nsw,hnsw,nsg,diskann,roar_graph,deg}. Tree-based methods perform well in low-dimensional datasets, but their search performance often degrades in high-dimensional spaces due to the curse of dimensionality. Hashing-based methods provide theoretical guarantees on search results, but in practice, it is difficult to maintain high search efficiency while achieving high accuracy. Quantization-based methods compress high-dimensional vectors into short quantization codes, significantly reducing the cost of distance computation. Graph-based methods achieve the best time-accuracy trade-off across various scenarios. Moreover, they can be combined with other methods to achieve better overall performance~\cite{symphonyQG, lsh_apg, hvs, efanna}. Since our method focuses on optimizing graph-based indexes, we next provide an overview of existing graph-based techniques.

Most state-of-the-art graph indexes are based on the Delaunay Graph~\cite{Delaunay} (DG) and the Relative Neighborhood Graph (RNG)~\cite{rng}. DG is the dual graph of the Voronoi diagram~\cite{aurenhammer}. In a DG, two vertices are connected by an edge if and only if there exists a circle passing through these two points, with no other points inside the circle. DG guarantees that for any given query $q$, a Greedy Search will be able to find $N_{1,q}$~\cite{hnsw}. However, DG becomes a complete graph in high-dimensional space~\cite{fanng}, resulting in search performance similar to brute-force search. RNG is a subgraph of DG. If an edge is the longest edge of any triangle formed by vertices in base data $X$, that edge is not included in the RNG. RNG guarantees that the average degree of each point is a small constant when points in $X$ are uniformly distributed, but it does not ensure the accuracy of the search. Later, Fu et al.~\cite{nsg} introduced the Monotonic Relative Neighborhood Graph (MRNG). MRNG preserves the property of constant average degree in RNG and guarantees that if query $q \in X$, a Greedy Search will find $N_{1,q}$ definitely. However, there is no theoretical guarantee for queries where $q \not\in X$. Subsequently, Fu et al.~\cite{ssg} extended MRNG to SSG. It provides some probabilistic guarantees for queries $q$ where $q \notin X$ under random distributions. Recently, Peng et al.~\cite{t_mg} introduced $\tau$-MG, which guarantees that if $\delta(N_{1,q},q) < \tau$, Greedy Search can find $N_{1,q}$. Due to the high graph construction overhead of the graph mentioned above, which makes them hard to use in practice, most state-of-the-art graph indexes are approximations of these algorithms. HNSW~\cite{hnsw} uses the RNG rule to sparsify the graph. It also uses a hierarchical structure, but the effect of this structure is limited in high-dimensional space~\cite{hnsw_h_useless, hnsw_h_useless2}. NSG, NSSG, and $\tau$-MNG are approximations of MRNG, SSG, and $\tau$-MG, respectively. They primarily consider the properties of these graphs in a local neighborhood around each point, thereby reducing the indexing construction overhead. However, these indexes tend to perform poorly when the query is far from the base data (e.g., certain ID queries or OOD queries).

To overcome the issues of hard queries, especially OOD queries, RobustVamana~\cite{ood_diskann} adds query vectors from other distributions into the graph and connects edges between the query and base points, where these query points serve as navigators. This partially mitigates the accuracy loss caused by OOD queries. However, these points also extend the search path, leading to only a small overall improvement. Chen et al.~\cite{roar_graph} proposed RoarGraph, which establishes a connection between the two distributions and avoids directly inserting query points into the graph by projecting them onto the base data. This approach achieves significantly better performance than existing graph indexes. 

There are also some other algorithms~\cite{learing_to_route,proba_route,Reinforcement_route,early_stop,ad_sampling,finger,iqan,hmann,hvs} that optimize the graph search process, such as iQAN~\cite{iqan}, which improves the performance and accuracy of ANNS by intra-query parallelism. LSH-APG~\cite{lsh_apg} and some other works~\cite{hmann,hvs} focus on solving the entry point selection problem. Additionally, some approaches~\cite{ad_sampling,finger,data_aware} aim to optimize the distance calculations during the search process.

\section{ANALYSIS}\label{analysis}
In this section, we mainly discuss the theoretical foundation for constructing graphs based on query distributions, along with preliminary experiments that support our design. 

To provide theoretical support, we first explore the relationship between the graph structure and accuracy for a given query. Following previous studies~\cite{scc,steiner_hardness}, we divide the search into two phases: (1) searching from the entry point to the vicinity of the query, and (2) searching within the vicinity of the query. Greedy Search can easily reach the second phase, but it often ends up retrieving only a subset of the nearest neighbors due to the lack of effective edges in the graph index ~\cite{steiner_hardness} (Figure \ref{search_path} illustrates an example of two search phases). Our experiments demonstrate that this conclusion holds for OOD queries as well. Figure \ref{recall} shows the proportion of queries with different recall levels when using HNSW as the graph index in four cross-modal datasets (the search list size is fixed at 100). For most queries, Greedy Search successfully reaches the query vicinity (i.e., recall@100 $\ne 0$). Therefore, we first focus on the case where the entry point is close to the query (Other cases will be discussed in Section \ref{sec::rf}). Subsequently, since Greedy Search only explores points closest to the query, we only need to consider the graph structure in the small region around the query. To better describe the area around the query, we first define the graph structure around a query~\cite{scc}, which consists of the query’s NNs and the edges associated with them:
\begin{figure}
    \centering
    \setlength{\abovecaptionskip}{0.1cm}
    \begin{subfigure}{0.3\linewidth}
		\centering
		\includegraphics[width=0.8\linewidth]{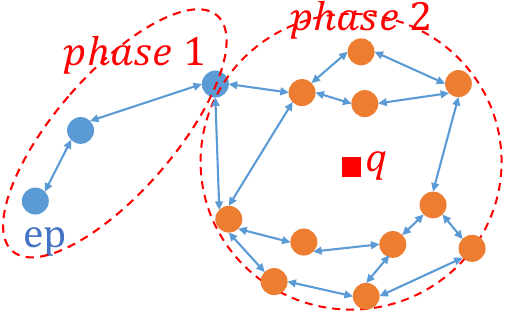}
		\caption{search phases}
		\label{search_path}
	\end{subfigure}
	\begin{subfigure}{0.33\linewidth}
		\centering
		\includegraphics[width=1\linewidth]{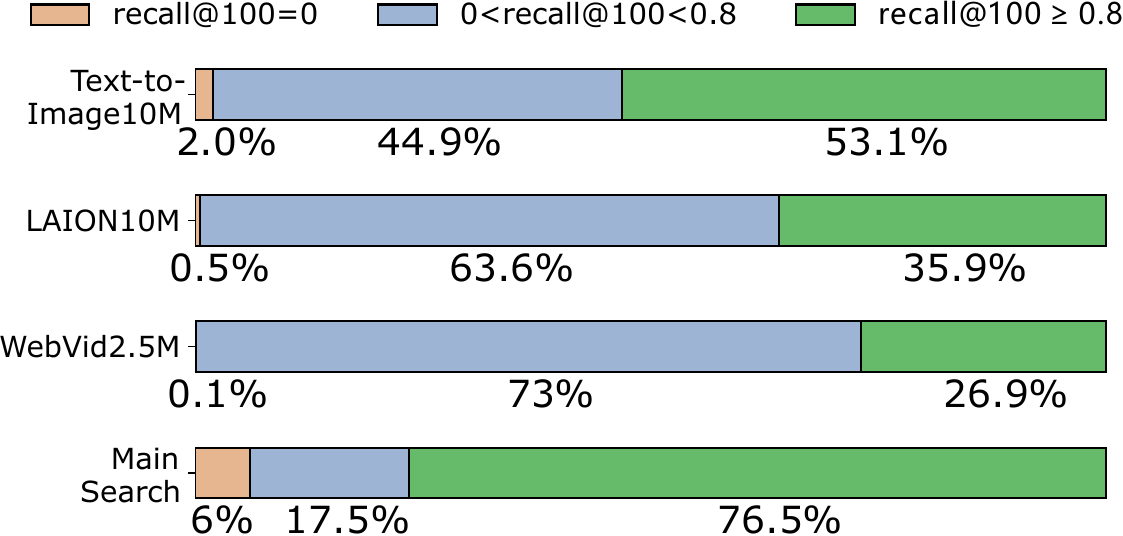}
		\caption{proportion of different queries}
		\label{recall}
	\end{subfigure}
    \caption{(a) The two phases of Greedy Search on the graph. (b) The proportion of queries with different recall levels. } 
    \Description{...}
    \label{search_phases}
    % \vspace{-1em}
\end{figure}

\begin{figure}
    \centering
    \setlength{\abovecaptionskip}{0.1cm}
	\begin{subfigure}{0.2\linewidth}
		\centering
		\includegraphics[width=0.7\linewidth]{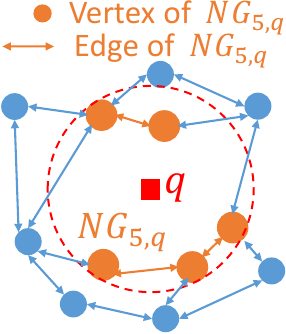}
		\caption{$NG_{5,q}$}
		\label{ng_example}
	\end{subfigure}
	\begin{subfigure}{0.4\linewidth}
		\centering
		\includegraphics[width=0.85\linewidth]{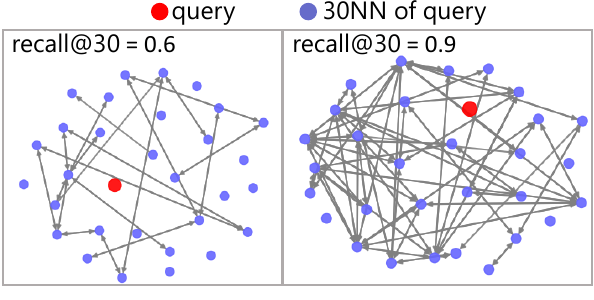}
		\caption{$NG_{30,q}$ in real-world dataset}
		\label{ng_real}
	\end{subfigure}
    \caption{(a) A toy example of $NG_{5,q}$. (b) $NG_{30,q}$ of HNSW's base layer with two different OOD queries in LAION10M.}
    \Description{...}
    % \vspace{-1em}
\end{figure}

\begin{definition}($S$-Neighboring Graph with query $q$ ($NG_{S,q}$)).
Consider a directed graph index $G = (V, E)$ and a query $q$, and define $V_{S,q} = \{N_{1,q},N_{2,q},...,N_{S,q}\}$. Then $NG_{S,q} = (V_{S,q}, E_{S,q})$ is the subgraph induced in $G$ by $V_{S,q}$, where $E_{S,q}$ is the edge set of the subgraph.

\end{definition}
Figure \ref{ng_example} shows an example of $NG_{5,q}$, where the orange vertices form $V_{5,q}$, and the orange edges form $E_{5,q}$. Figure \ref{ng_real} illustrates $NG_{30,q}$ for two OOD queries on the LAION10M~\cite{laion} dataset, using an index constructed with HNSW's base layer. To facilitate visualization, we applied Multidimensional Scaling (MDS)~\cite{mds} to project the high-dimensional data to two dimensions. The recall@30 shown in the figure represents the results of Greedy Search with a randomly selected entry point and a search list size of 30. It can be observed that for queries with poor accuracy, $NG_{30,q}$ exhibits weaker connectivity and contains numerous isolated points. 

We then examine the relationship between $NG_{S,q}$ and query accuracy from two perspectives: (1) A high-quality $NG_{S,q}$ guarantees high query accuracy; (2) A low-quality $NG_{S,q}$ can easily lead to a significant drop in accuracy. \textbf{For the first perspective}, we introduce the following theorem to show that if base points $u$ and $v$ are connected in $NG_{S,q}$, then Greedy Search starting from $u$ is guaranteed to reach $v$ when the search list size $L\ge S$. All the proof in this paper and more details can be found in Appendix B \footnote{\href{https://github.com/yuhuifishash/NGFix/blob/main/appendix.pdf}{https://github.com/yuhuifishash/NGFix/blob/main/appendix.pdf}}.
\begin{theorem}\label{ng}
    For a directed graph index $G = (V, E)$ and a query $q$. If $N_{i,q}$ can reach $N_{j,q}$ in $NG_{S,q}$ ($1 \le i,j \le S$). The Algorithm \ref{beamsearch} will always visit $N_{j,q}$ when query $=q$, $ep = N_{i,q}$ and $L \ge S $.
\end{theorem}

\begin{figure}
    \centering
    \setlength{\abovecaptionskip}{0.1cm}
    \begin{subfigure}{0.3\linewidth}
		\centering
		\includegraphics[width=1\linewidth]{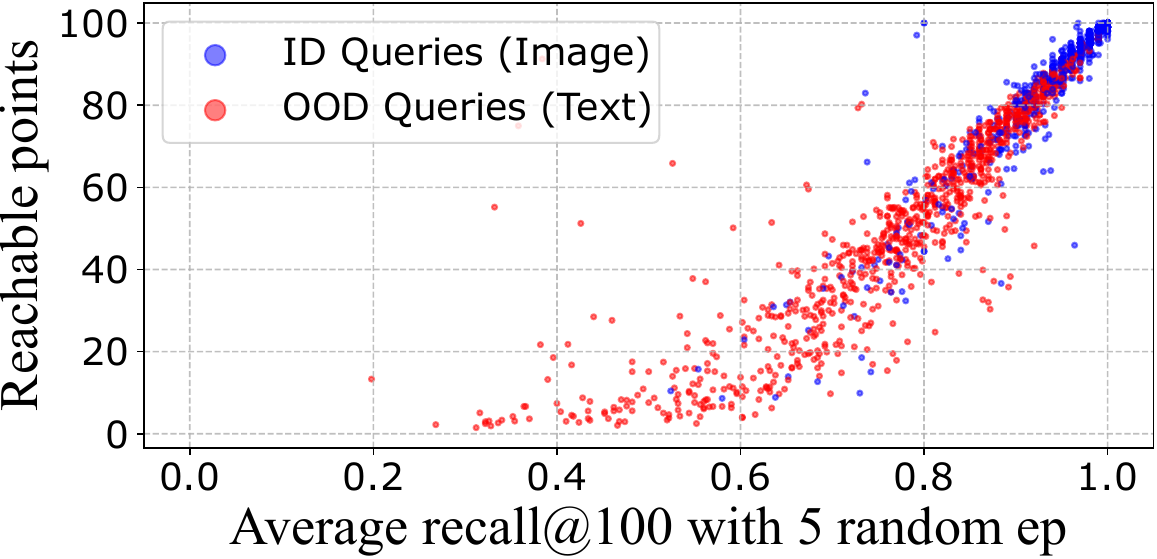}
		\caption{Correlation on LAION10M}
		\label{connective_corr}
	\end{subfigure}
	\begin{subfigure}{0.3\linewidth}
		\centering
		\includegraphics[width=1\linewidth]{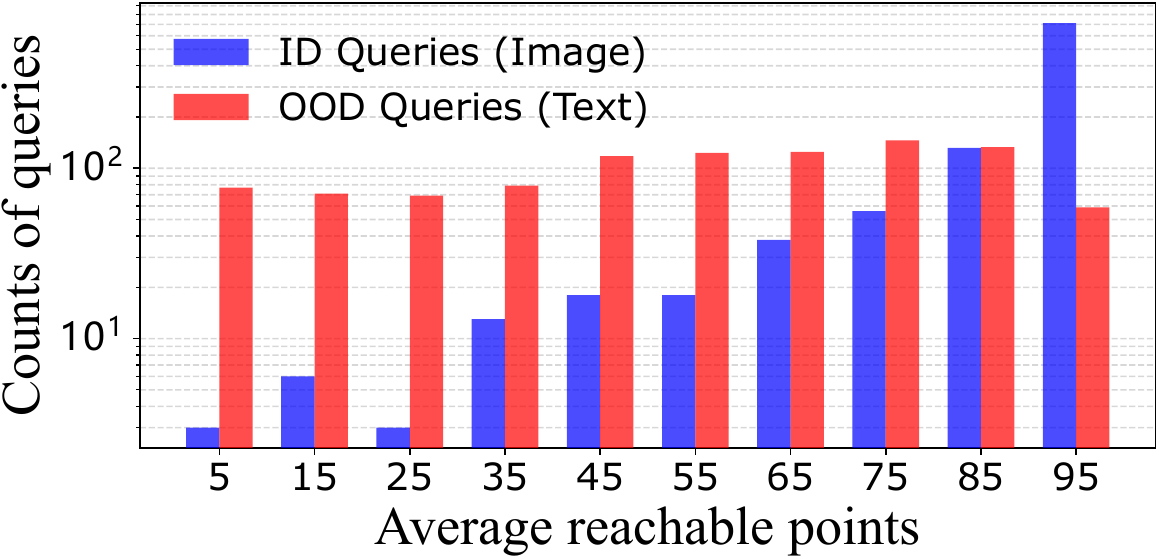}
		\caption{Counts on LAION10M}
		\label{connective_laion}
	\end{subfigure}
    \caption{Randomly sampled 1000 ID queries and 1000 OOD queries in LAION10M. The graph index is the base layer of HNSW. (a) The correlation between query's recall@100 with a search list size of 100 and its average number of reachable points in $NG_{100,q}$. (b) Counts of queries with different average numbers of reachable points in $NG_{100,q}$. } 
    \Description{...}
    \label{connective_example}
    % \vspace{-1em}
\end{figure}

Theorem \ref{ng} demonstrates that a $NG_{S,q}$ with strong connectivity can guarantee search accuracy. Furthermore, \textbf{for the second perspective}, our experiments show that poor connectivity in the $NG_{S,q}$ significantly degrades search accuracy. We measure the connectivity of $NG_{S,q}$ by calculating the average number of points that can be reached starting from a randomly chosen point in $NG_{S,q}$. Figure \ref{connective_corr} shows the correlation between query accuracy and the average number of reachable points in $NG_{100,q}$, demonstrating that insufficient connectivity in the $NG_{S,q}$ can lead to a substantial drop in search accuracy. Moreover, it can be observed that a small number of queries have a strongly connected $NG_{S,q}$, but their average recall@100 is still below 1. This is because for these queries, if the search starts from certain entry points, it may fail to reach the neighborhood of the query (i.e., fail to enter the second phase of the search).

\textbf{Connectivity of $NG_{S,q}$ in Real-World Datasets.} Figure \ref{connective_laion} shows the connectivity of $NG_{100,q}$ in LAION10M. We observe that the connectivity of their $NG_{100,q}$ varies significantly for both ID and OOD queries. Although the connectivity of $NG_{100,q}$ for OOD queries is poorer compared to that of ID queries in general, there are still some special cases: approximately 30\% of OOD queries have a highly connected $NG_{100,q}$, while about 10\% of ID queries have an $NG_{100,q}$ with very few edges. This means that when using queries to guide graph construction, we should focus on the hard queries with low-quality $NG_{S,q}$.

\textbf{Reason for Using Historical Queries.} Although the graph structure around the query impacts the query accuracy, the following theorem we propose demonstrates that for any query and $S$, ensuring strong connectivity in $NG_{S,q}$ is impractical.

\begin{theorem}\label{dg}
Given a base data set $X$ and the DG constructed from the points in $X$. If any edge of DG is removed, there exists a query $q$ such that $NG_{2,q}$ contains only two isolated nodes.
\end{theorem}

Since DG becomes a complete graph in high-dimensional spaces~\cite{fanng}, ensuring strong connectivity for the $NG_{S,q}$ of all queries is impractical in the real world. Therefore, the method we propose below leverages the historical queries to achieve strong connectivity for the $NG_{S,q}$ in regions where queries are densely distributed.
 
\section{DESIGN}

\begin{figure*}\centering
    \setlength{\abovecaptionskip}{0.1cm}
	\includegraphics[width=5.4in]{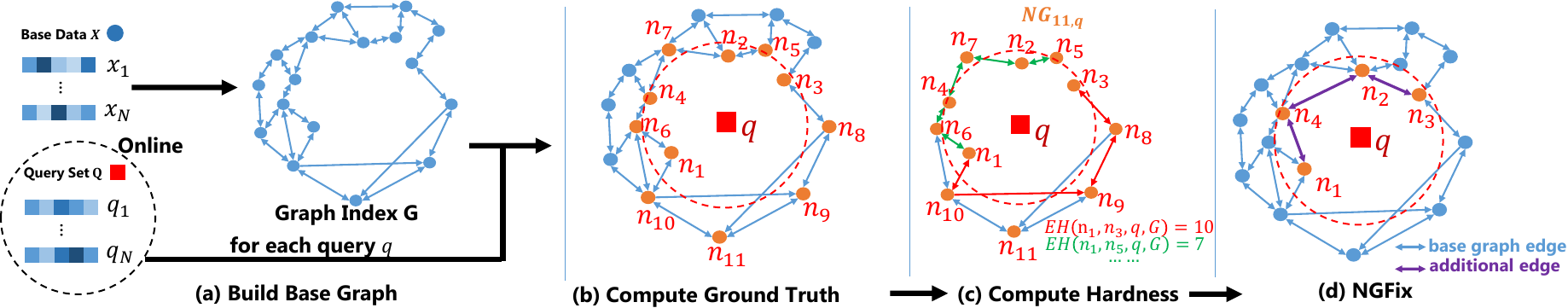}
	\caption{An overview of Neighboring Graph Defects Fixing with $N_q = 4, K_h = 4$.}
    \Description{..}
    \label{Overview}
    % \vspace{-1em}
\end{figure*}

The analysis in Section \ref{analysis} can be summarized as follows: (1) The difficulty of a query is largely influenced by the quality of the graph structure around it, and the difficulty varies significantly across queries. So we first propose EH to measure the quality of the graph structure around a query in this section. (2) We need to utilize historical queries to enhance graph connectivity in regions where queries are densely distributed. So we then propose NGFix, using EH in combination with historical queries to guide us in fixing the defective areas of the graph. Figure \ref{Overview} illustrates the overview of NGFix. We begin by constructing a base graph (Figure \ref{Overview}a). For each query $q$, we compute its exact $K$NN (Figure \ref{Overview}b) and then calculate EH for $q$ (Figure \ref{Overview}c). Finally, we add additional edges in the vicinity of $q$ to fix the defective regions of the graph (Figure \ref{Overview}d). (3) There still exist a few queries for which the search stops in the first phase and cannot reach their vicinity. So we propose RFix to enhance the navigability of some points. 

\subsection{Preprocess}\label{sec:preprocess}
First, we need to construct any type of graph index as a base graph, such as HNSW, NSG, etc. Then, for each query, we need to compute its exact $k$NN. We have the following methods for calculating $k$NN:
\begin{itemize}[leftmargin=0.4cm, itemindent=0cm]
    \item Collect the queries and process them after accumulating a certain amount (e.g., 512), and then transform the computation of exact $k$NN into matrix multiplication, and accelerate it using matrix multiplication optimizations~\cite{yael}. 
    \item Utilize the previously constructed graph index to perform Greedy Search, but increase the search list size to improve the accuracy of approximate $k$NN. This method is significantly faster than brute-force computation, but may slightly affect the ground truth quality of the queries. The results of experiments in Section \ref{sec:appro_gt} demonstrate that this method has little impact on the final graph's quality.
\end{itemize}

\subsection{Escape Hardness}

\begin{figure}
    \centering
    \setlength{\abovecaptionskip}{0.1cm}
    \includegraphics[width=0.7\linewidth]{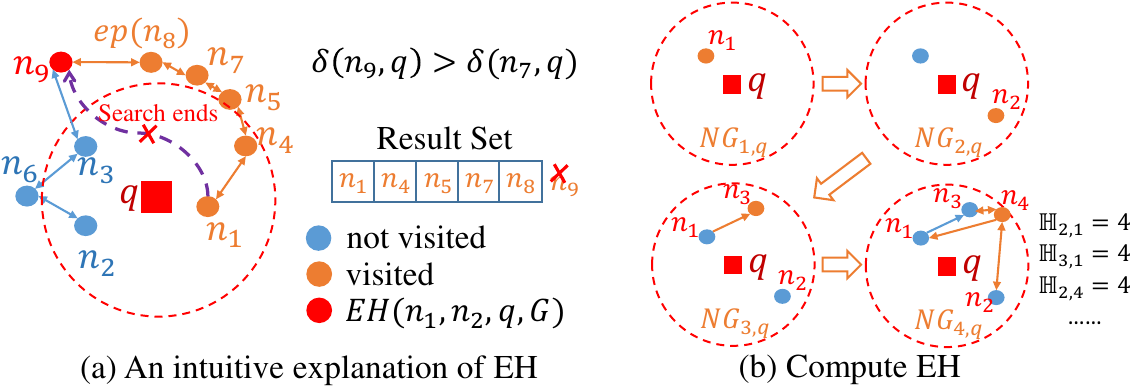}
    \caption{Example of Escape Hardness.}
    \Description{..}
    \label{eh_example}
    % \vspace{-1em}
\end{figure}

To better measure the quality of the graph structure around a query, we introduce Escape Hardness. It focuses on measuring the difficulty of reaching one point from another around the query using Greedy Search.

\begin{definition}(Escape Hardness (EH)).
    Given a directed graph index $G = (V, E)$ and a query $q$, let $P(u,v,G)$ denotes the set of all paths from vertex $u$ to vertex $v$ in graph $G$ and $F_{u,q}$ denote the point $u$ is the $F_{u,q}$-th NN of $q$, the Escape Hardness of query $q$ from vertex $u$ to vertex $v$ is defined as:
    $$ EH(u,v,q,G) = \underset {p \in P(u,v,G)}{min} \underset {x \in p}{max} F_{x,q}$$
    and we define the Escape Hardness Matrix $\mathbb{H} \in \{1,2,...,|X|\}^{N_q \times N_q}$ where $\mathbb{H}_{i,j} = EH(N_{i,q}, N_{j,q}, q, G)$ and $N_q$ represents the number of nearest neighbors considered for query $q$. 
\end{definition}

The definition of $EH(u,v,q,G)$ illustrates the hardness of searching from a point $u$ to reach $v$ using Greedy Search. The validity of EH is primarily supported by the following corollary of Theorem \ref{ng}. EH establishes an upper bound on the search list size required to successfully reach a point from another point around the query.

\begin{Corollary}
    \label{kh_corollary}
    For a directed graph index $G = (V, E)$ and a query $q$. the Algorithm \ref{beamsearch} will always visit $N_{j,q}$ when query $=q$, $ep = N_{i,q}$ and $L \ge EH(N_{i,q},N_{j,q},q,G) $.
\end{Corollary}

\textbf{Example.} Figure \ref{Overview}c is an example for EH: For simplicity, we use $n_i$ to represent $N_{i, q}$ in the figure. For $EH(n_1, n_3, q, G)$, without considering the detour through $n_6$, there are mainly two paths from $n_1$ to $n_3$: $p_1 = \{n_1, n_{10}, n_9, n_8, n_3\}$ and ${max}_{x \in p_1} F_{x,q} = 10$, $p_2 = \{n_1, n_{10}, n_{11}, n_9, n_8, n_3\}$ and ${max}_{x \in p_2} F_{x,q}=11$. Since ${max}_{x \in p_1} F_{x,q}$ is smaller, $EH(n_1, n_3, q, G) = 10$, and $\mathbb{H}_{1,3} = 10$. \textbf{Intuition.} In Figure \ref{eh_example}a, we take $n_8$ as the entry point and aim to recall the top-2 NN of $q$ using a result set of size $5$ (i.e., $L = 5$ in Algorithm \ref{beamsearch}). After the search visits $n_1$, the result set becomes full. Since $n_9$ is far from the query, it will be removed from the result set. Then the greedy search converges prematurely. It will not go back to $n_9$ and expand its neighbors, resulting in recall@2 = 50\%. Here, EH measures the difficulty of reaching $n_2$ from $n_1$ along the search path. Specifically, $EH(n_1, n_2, q, G) = 9$, meaning the critical point ($n_9$) hindering the search ranks 9th in terms of distance to $q$. In this example, increasing $L$ to $6$ would allow the search to reach $n_2$. However, it is generally difficult to determine the exact $L$ needed to ensure a certain point is visited, so EH provides an upper bound on $L$.

To compute EH, we first present the following theorem:
\begin{theorem}\label{eh_calculate}
    If points $N_{i,q}$ and $N_{j,q}$ satisfy the condition that $N_{i,q}$ cannot reach $N_{j,q}$ in $NG_{S-1,q}$ but can reach $N_{j,q}$ in $NG_{S,q}$, then $EH(N_{i,q},N_{j,q},q,G) = S$ and $\mathbb{H}_{i,j}=S$.
\end{theorem}

\begin{algorithm}
\small
\setstretch{0.8} 
\caption{Compute Hardness}\label{EH_Compute}
\KwIn{Query $q$, $NG_{MaxS,q} = (V_{MaxS,q}, E_{MaxS,q})$, $N_q$}
\KwOut{The Escape Hardness Matrix $\mathbb{H}$}
Initialize a matrix $f \in \{0,1\}^{MaxS\times MaxS}$ with all elements set to $0$\\
Set all elements of $\mathbb{H}$ to $\infty$\\
Let $f_{i,j} \leftarrow 1$ if $(i,j) \in E_{MaxS,q}$ or $i = j$ for all $1\le i,j \le N_q$\\
Let $\mathbb{H}_{i,j}\leftarrow max(i,j)$ if $f_{i,j}=1$ for all $1\le i,j \le N_q$\\
\For(\tcp*[f]{add points to subgraph.}){$h \leftarrow 1$ \KwTo $MaxS$}{
    \For{$i \leftarrow 1$ \KwTo $MaxS$}{
        $C \leftarrow \emptyset$ \tcp*[f]{Save the connectivity-changed points.} \\
        \For(\tcp*[f]{Floyd-Algorithm.}){$j \leftarrow 1$ \KwTo $MaxS$}{
            $f_{i,j} \leftarrow f_{i,j}$ $or$ $(f_{i,h}$ $and$ $f_{h,j})$ \\
            \textbf{if} {$f_{i,j}$ changed}  \textbf{then} $C \leftarrow C \cup \{j\}$\\
            
        }
        \If{$i \le N_q$}{
            \For{$u \in C$ and $u \le N_q$}{
                $\mathbb{H}_{i,u} = max(i,u,h)$
            }
        }
    }
}
\Return $\mathbb{H}$
\end{algorithm}

Due to THEOREM \ref{eh_calculate}, when computing EH, we start with $S=1$ and incrementally consider $NG_{S,q}$ in ascending order of $S$. For the subgraph $NG_{S-1,q}$, we add the point $N_{S,q}$ and its associated edges to construct $NG_{S,q}$. If the connectivity from points $N_{i,q}$ to $N_{j,q}$ changes after adding point $N_{S,q}$, then $\mathbb{H}_{i,j}=S$. The connectivity updates after each addition are computed using a modified Floyd-Warshall algorithm~\cite{floyd}. The calculation of EH is finished when $N_{i,q}$ and $N_{j,q}$ are reachable from each other for any $1 \le i,j \le N_q$. We denote the value of $S$ at the time the computation is completed as $MaxS$, also meaning that the subgraph at this point is $NG_{MaxS,q}$. However, based on the previous analysis, the connectivity of the graph around hard queries tends to be poor, often resulting in a significantly large value of $MaxS$. Therefore, to efficiently compute EH, we limit $MaxS$ to a constant multiple of $N_q$ (e.g., $MaxS \le 5N_q$). If $N_{i,q}$ and $N_{j,q}$ are still not connected at this point, we consider $\mathbb{H}_{i,j}$ to be infinite. Figure \ref{eh_example}b shows an example of computing EH: After adding $n_1$ and $n_2$, none of the nodes can reach each other, so we cannot compute any EH now. When $n_3$ is added, $n_1$ becomes able to reach $n_3$, so we set $\mathbb{H}_{1,3}=3$. After adding $n_4$, $n_2$ can reach $n_1$ through $n_4$, so $\mathbb{H}_{2,1}=4$.
For other node pairs that also become reachable via $n_4$, their EH values are also set to $4$. 

The details of the algorithm are shown in Algorithm \ref{EH_Compute}. To accelerate the computation, we directly set $MaxS$ to $5N_q$ and preprocess $NG_{MaxS,q}$ before calculation. The matrix $f$ represents the transitive closure, and $f_{i,j}$ indicates whether $N_{i,q}$ can reach $N_{j,q}$ (line 1). Then we set the initial values of $f$ and $\mathbb{H}$ based on the edges in $NG_{MaxS,q}$ (lines 2-4). The loop in line 5 adds a new point in each iteration. After adding points, we perform the Floyd algorithm to compute the connectivity of the new graph and store the points where connectivity changes (lines 6-10). Finally, we update the matrix $\mathbb{H}$ based on the changes (lines 11-13). In our implementation, since the matrix $f$ is a boolean matrix, we use bitset to store $f$ and speed up the Floyd algorithm. 

\textbf{Comparison to $Steiner$-hardness ~\cite{steiner_hardness}.} Recently, $Steiner$-hardness effectively estimates the query difficulty with high accuracy. However, its focus differs from that of EH: (1) EH provides a finer-grained measure of the difficulty of reaching each point from any other in a query (represented as a matrix), while $Steiner$-hardness gives an overall difficulty score (single value). (2) Steiner Hardness focuses on estimating the computational cost required to achieve a target accuracy (e.g., recall@100 = 0.9), whereas EH aims to provide theoretical support for graph construction by measuring the upper bound of the search list size required for greedy search to reach a target from a starting point.

\subsection{Neighboring Graph Defects Fixing}
\begin{figure}
    \centering
    \setlength{\abovecaptionskip}{0.1cm}
	\begin{subfigure}{0.2\linewidth}
		\centering
		\includegraphics[width=1\linewidth]{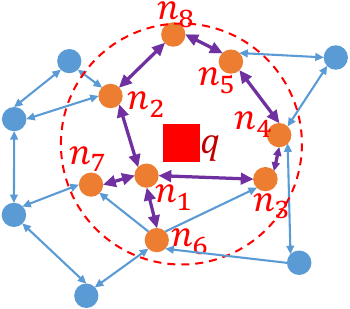}
		\caption{RNG}
		\label{defects_fix_rng}
	\end{subfigure}
	\begin{subfigure}{0.2\linewidth}
		\centering
		\includegraphics[width=1\linewidth]{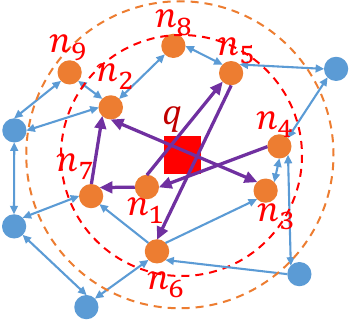}
		\caption{Random}
		\label{defects_fix_random}
	\end{subfigure}
    \begin{subfigure}{0.2\linewidth}
		\centering
		\includegraphics[width=1\linewidth]{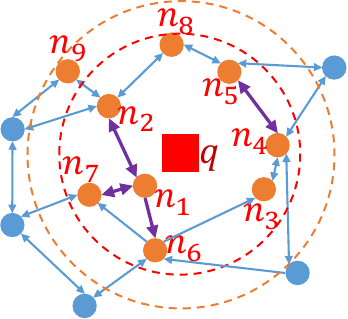}
		\caption{NGFix}
		\label{defects_fix_example}
	\end{subfigure}
    \caption{The results of different methods for fixing the $NG_{8,q}$, where blue edges represent the edges of the base graph and purple edges represent the additional edges we add. (a) Reconstructing the RNG graph in the $NG_{8,q}$. (b) Randomly adding edges until any two points in the $NG_{8,q}$ are $K_h$-reachable ($K_h=9$). (c) NGFix with $N_q=8$ and $K_h = 9$.}
    \Description{...}
    \label{defects_fix_different_example}
    % \vspace{-1em}
\end{figure}

\textbf{Using Escape Hardness to guide graph construction. }Since Escape Hardness describes the hardness of Greedy Search traversing from one point to another around a query, our key idea is to \textbf{ensure that the Escape Hardness between all point pairs in the small region surrounding each historical query is low} (i.e., fix the defects in the graph). We first define a threshold $K_h(K_h \ge N_q)$ to determine whether it is hard to travel from one point to another.

\begin{definition}($K_h$-reachable).
Given a directed graph index $G = (V, E)$ and a query $q$, if point $u$ and $v$ satisfy $EH(u,v,q,G) \le K_h$, we consider $u$ to be $K_h$-reachable to $v$. Then we define the $K_h$-reachable Matrix $\mathbb{T} \in \{0,1\}^{N_q \times N_q}$ as:
$\mathbb{T}_{i,j} = \begin{cases}
1 & \mathbb{H}_{i,j} \le K_h \\
0 & \mathbb{H}_{i,j} > K_h \\
\end{cases}$
\end{definition}

According to Corollary \ref{kh_corollary}, ensuring low Escape Hardness between point pairs guarantees the accuracy of Greedy Search. Therefore, our next step is to develop an effective method to fix the defects in the graph.

\begin{sloppypar}
\textbf{Simple Solutions and Limitations.} (1) Reconstruct RNG within the $NG_{S,q}$ for each historical query $q$ and overlay it with the original graph. Figure \ref{defects_fix_rng} illustrates an example of constructing the RNG graph. This method provides high-quality neighbors for each point in $NG_{S,q}$. However, it connects many edges for individual queries, which increases the computational burden during search and enlarges the memory footprint of the graph index. (2) Randomly selecting points $N_{i,q}$ and $N_{j,q}$ with $\mathbb{T}_{ij}=0$ in $NG_{S,q}$ and adding edges until each point is $K_h$-reachable to any other points in $NG_{S,q}$ (Figure \ref{defects_fix_random}). However, this method results in disordered connections, and the points in $NG_{S,q}$ are not connected to their actual neighbors.
\end{sloppypar}

\begin{algorithm}
\small
\setstretch{0.8} 
\caption{Neighboring Graph Defects Fixing}\label{defects_fix}
\KwIn{Query $q$, Graph Index $G = (V,E,E_{EX})$, $N_q$, degree limitation of extra edges $M_{EX}$, $\mathbb{T}$, $\mathbb{H}$}
\KwOut{New Graph Index $G_{EX}$}
$EC \leftarrow \emptyset$ \tcp*[f]{Edge Candidates} \\ 
\For{$(i,j) \in \{(i,j)|\mathbb{T}_{i,j} = 0, 1 \le i,j \le N_q\}$}{
    $EC \leftarrow EC \cup \{(i,j)\} $
}
Sort $(i,j) \in EC$ in ascending order by $\delta(N_{i,q},N_{j,q})$\\
\While{$EC \ne \emptyset$}{
    $(s,t) \leftarrow $ pop the first element of $EC$\\
    \textbf{if} $\mathbb{T}_{s,t} \ne 0$ \textbf{then} continue;\\
    $\mathbb{T}_{s,t} \leftarrow 1$\\
    \uIf(\tcp*[f]{Link new edge}){$|ExtraNeighbors(N_{s,q})$$|$ $< M_{EX}$ }{
        $E_{EX} \leftarrow E_{EX} \cup \{(N_{s,q},N_{t,q},\mathbb{H}_{s,t})\}$ \\
    }\Else( \tcp*[f]{Prune Edges}){
        choose $(N_{s,q},v,h) \in E_{EX}$ with minimum $h$\\
        \uIf{$h < \mathbb{H}_{s,t}$}{
            $E_{EX} \leftarrow E_{EX}\setminus\{(N_{s,q},v,h)\}$\\
            $E_{EX} \leftarrow E_{EX} \cup \{(N_{s,q},N_{t,q},\mathbb{H}_{s,t})\}$\\
        }
    }
    \For{$i \leftarrow 1$ \KwTo $N_q$}{
        \For{$j \leftarrow 1$ \KwTo $N_q$}{
            $\mathbb{T}_{i,j} \leftarrow \mathbb{T}_{i,j}$ $or$ $(\mathbb{T}_{i,s}$ $and$ $\mathbb{T}_{t,j})$\tcp*[f]{update $\mathbb{T}$.}\\
        }
    }
} 
\Return $G_{EX} = (V, E, E_{EX})$

\end{algorithm}

\textbf{NGFix.} After considering both the graph's neighborhood properties and the number of additional edges, we choose to use the idea of the minimum spanning tree (MST)~\cite{mst}, and MST is a subset of RNG~\cite{rng}. We consider each edge $(N_{i,q},N_{j,q})$ in increasing order of weight $\delta(N_{i,q},N_{j,q})$. If $\mathbb{T}_{i,j}=0$, we add that edge to the graph. Additionally, adding a new edge will make some points $K_h$-reachable to others, so we update $\mathbb{T}$ after each edge is added to the graph. This process is repeated until all elements of $\mathbb{T}$ are equal to $1$. Figure \ref{defects_fix_example} and \ref{Overview}d show the final result of NGFix in a toy example. The details of the algorithm are shown in Algorithm \ref{defects_fix}. We represent a graph index as $G=(V,E,E_{EX})$, where $E$ denotes the edges of the base graph, and $E_{EX}$ represents the additional edges we add. The main part of the algorithm is a loop that considers each edge in ascending order of distance. In each iteration, we first check whether the edge needs to be added (lines 6-8). Then, we set an additional out-degree limit $M_{EX}$ to prevent any point from having too many extra edges (lines 10-11). When a point exceeds this limit, we prioritize pruning the edges with lower EH (lines 13-16). Finally, we update $\mathbb{T}$. The key idea is that for all $1\le i,j \le N_q$, if $N_{i,q}$ is $K_h$-reachable to $N_{s,q}$ and $N_{t,q}$ is $K_h$-reachable to $N_{j,q}$, then after connecting $N_{s,q}$ and $N_{t,q}$, $N_{i,q}$ will become $K_h$-reachable to $N_{j,q}$ (lines 17-19).

\textbf{Analysis.} The following theorem provides the maximum number of edges that NGFix may add for a given query.
\begin{theorem}\label{complexity}
    Given a query $q$, Algorithm \ref{defects_fix} will add at most $2(N_q-1)$ extra directed edges. 
\end{theorem}

Based on Theorem \ref{complexity}, the loop in Algorithm \ref{defects_fix} will execute at most $O(N_q)$ times before terminating. Each iteration of the loop needs to update $\mathbb{T}$, which takes $O(N_q^2)$ time, so the time complexity of Algorithm \ref{defects_fix} is $O(N_q^3)$. For the calculation of EH (Algorithm 2), we need to add the $MaxS$ nearest neighbors of the query sequentially, and each addition requires $O(MaxS^2)$ time to compute connectivity. Therefore, the time complexity of EH calculation is $O(MaxS^3)$. Since we limit $MaxS$ to at most a constant multiple of $N_q$, the time complexity of our algorithm for a single query is $O(P + N_q^3)$, where $P$ represents the time complexity of preprocessing (Section \ref{sec:preprocess}). And each query will at most increase the size of the graph index by $O(N_q)$. Since $N_q$ is generally set to a small value (e.g., 100), the main time overhead of our algorithm lies in the preprocessing step (i.e., computing the ground truth of each historical query).

\subsection{Reachability Fixing}\label{sec::rf}

Previously, we focused on cases where the search successfully reaches the vicinity of the query. However, for a small subset of queries, Greedy Search fails to do so and stops before approaching the query. In the following, we discuss how to address situations where the search terminates far from the query. When this occurs, it indicates that the current point lacks outgoing edges leading toward the query. Recall that most existing graph index construction strategies approximate the RNG or its variants\cite{nsg,hnsw,roar_graph,t_mg,ssg}. Their approximation process typically involves two steps: (1) For a given point, treat it as a query and use Greedy Search to retrieve nearby points, typically with a result set size ranging from 100 to 2000. (2) Apply an edge pruning strategy (e.g., MRNG) to the result set to select the true neighbors. However, since candidate neighbors are selected from a relatively small greedy search result set, the retrieved points may be clustered in a single direction. As a result, valuable neighbors in other directions may be overlooked. If the query lies in one of these neglected directions, the Greedy Search is likely to stop early without finding closer points to the query.

According to the analysis above, we propose Reachability Fixing (RFix). The key idea of RFix is to \textbf{expand the pruning candidate set of the $ANN_{1,q}$'s neighbors for searches that fail to reach the vicinity of the historical query}, where $ANN_{1,q}$ refers to the approximate nearest neighbor returned by the Greedy Search. The details are shown in Algorithm \ref{rf} and the steps are as follows: (1) \textbf{Select the centroid of the base data as the entry point.} For a historical query, it is difficult to guarantee that Greedy Search starting from every entry point will reach the vicinity of the query. Therefore, we fix the entry point as the centroid, which also helps reduce the length of the search path. (2) \textbf{Expand the candidate neighbor set of $ANN_{1,q}$.} First, we perform a Greedy Search to retrieve the $ANN_{1,q}$ to the historical query (line 2). If the search did not reach the vicinity of the query (line 3), we expand the candidate neighbors of $ANN_{1,q}$. First, we collect all points $u$ that satisfy $\delta(u, q) < \delta(ANN_{1,q}, q)$ as the extended candidate neighbor set (line 4). Then, we apply the RNG pruning strategy to this set to ensure that the angle between any two connected edges is greater than $60^{\circ}$ (lines 5-9), which disperses the edges in different directions and enhances the navigational performance of the graph. Since these edges provide important paths to reach the query, their EH is set to infinity (line 9). Finally, we add these edges to the graph using the same EH pruning strategy as in NGFix (line 10). For the computation of extended candidate neighbor set (line 4), brute-force searching would be time-consuming. Therefore, we use a greedy search with a larger search list size and check all the points visited during the search process to replace the brute-force search.

\begin{algorithm}[t]
\small
\setstretch{0.8}
\caption{Reachability Fixing}\label{rf}
\KwIn{Query $q$, Graph index $G = (V,E,E_{EX})$, $N_q$}
\KwOut{New Graph Index $G_{EX}$}
$ep \leftarrow $ centroid of base data \\
$ANN_{1,q} \leftarrow GreedySearch(G, q, 1, ep, N_q)$ \tcp*[f]{Algorithm \ref{beamsearch}}\\

\uIf(\tcp*[f]{can not reach query}){$\delta(ANN_{1,q},q) > \delta(N_{N_q,q},q)$}{
    $S \leftarrow \{u|u\in X, \delta(u, q) < \delta(ANN_{1,q}, q)\}$\\
    Sort points in $S$ in ascending order of their distance to $ANN_{1,q}$.\\
    $Candidates \leftarrow \emptyset$ \tcp*[f]{candidate neighbor set}\\
    \For(\tcp*[f]{MRNG-based pruning}){ $v \in S$ }{
        \uIf{$\forall r \in Candidates$, $\delta(v, r) > \delta(ANN_{1,q}, v)$}{
            $Candidates \leftarrow Candidates \cup \{(ANN_{1,q}, v, inf)\}$
        }
    }
    $E_{EX} \leftarrow E_{EX} \cup Candidates$ \\ \tcp*[f]{Use the same EH pruning strategy as in lines 11-15 of Algorithm \ref{defects_fix}.}
}
\Return $G_{EX} = (V, E, E_{EX})$
\end{algorithm}

Since a single RFix does not guarantee that the search will reach the vicinity of the query, we repeat the RFix until the search can reach the query's vicinity or the edge degree limit for that point is reached. As only a small number of historical queries encounter issues where the search cannot reach the query's vicinity, RFix does not significantly increase the graph construction time or index size. After applying both NGFix and RFix, we have the following theorem.

\begin{theorem}\label{accuracy}
    Let $T$ denote the set of historical queries. Without pruning the edges, our method ensures that when $q \in T$ and $k \le N_q$, the accuracy of GreedySearch($G$, $q$, $k$, centroid, $K_h$) is 100\% (Algorithm \ref{beamsearch}).
\end{theorem}

Recall that MRNG guarantees the accuracy of Greedy Search when $q \in X$ (base data), experiments show that the search accuracy remains high when $q$ is close to $X$~\cite{nsg}. Our algorithm serves a similar purpose as MRNG, but the key difference is that we construct the graph on the query distribution, whereas MRNG constructs the graph on the base data distribution. The reasons for setting an out-degree limit for each point are as follows: (1) to prevent the graph index from becoming too large; (2) an excessive number of edges will benefit historical queries, but it does not necessarily improve the performance of other queries and may increase the computational burden during the search process. 

\subsection{Index Maintenance}

We have introduced how to dynamically update the graph structure based on queries. However, dynamic addition and deletion of base data is also a critical issue in real-world applications~\cite{spfresh, fresh-diskann}. Next, we discuss how the graph index handles insertion and deletion operations on the base data. 

\subsubsection{Insertion} \label{sec:insertion}
To support dynamic insertion, we require a base graph structure that allows incremental updates (e.g., HNSW). When a new data point arrives, it is first inserted into the base graph following its insertion algorithm. However, after a large number of insertions, the extra edges linked by NGFix and RFix may have limited impact on newly inserted points, which leads to a degradation in search performance. To address this issue, we propose a partial rebuilding strategy with the following steps: (1) For each node, randomly remove a proportion $r$ (e.g., 20\%) of its extra outgoing edges (will not remove the edge of base graph), and reset the EH values of the remaining edges to zero. This is because the previously estimated hardness may no longer reflect the current graph structure accurately. (2) Randomly select $r|T|$ historical queries from the set of historical queries $T$ and apply NGFix and RFix on the updated graph. A smaller $r$ leads to faster reconstruction but may result in lower graph quality. We provide a detailed analysis of the trade-offs in Section \ref{sec:index_maintenance}.

\subsubsection{Deletion}
The most commonly used method for deleting points is lazy deletion~\cite{hnsw,roar_graph,t_mg}, where the deleted points are retained for navigation during the search process but are excluded from the final result set. This method is simple and efficient. However, when a large number of points are deleted, the deleted points in the graph can significantly prolong the search paths, resulting in degraded search performance. To address this issue, we need to completely remove the point and its associated edges from the graph. This introduces two main challenges: (1) We need to identify and remove all incoming edges to the deleted point. A trivial solution is to store all incoming edges for each point in memory, but this significantly increases memory consumption. (2) The deletion of the point may damage the structure around it, resulting in reduced connectivity in its neighborhood.

For the first challenge, we adopt a lazy deletion strategy for a small number of deletions. Once the number of deleted points exceeds a certain threshold (e.g., 1\% of the base data), we perform a full traversal of the graph index to remove all deleted points along with their incoming edges. For the second challenge, recall that NGFix is designed to fix defective graph structures in the vicinity of queries. Thus, we can simply treat the deleted point as a query and execute NGFix. The connectivity degradation resulting from the point deletion will be mitigated by applying NGFix.

\section{EXPERIMENTS}
\begin{table}
\centering
\begin{tabular}{ccccccc}
\hline
DataSet       & $|X|$ & $|Q|$ & $|T|$ & $d$ & Distance & Type                     \\ \hline
Text-to-Image10M & 10M   & 10K   & 10M     & 200 & InnerProduct       & Text, Image \\ \hline
LAION10M              & 10M      & 10K      & 10M        & 512    & Cosine          &        Text, Image                  \\ \hline
WebVid2.5M              & 2.5M      & 10K      & 2.5M        & 512    & Cosine         &  Text, Video                       \\ \hline
MainSearch             &  11.2M     &  50K     &  1M       & 256    & InnerProduct         & Text, Image                         \\ \hline
SIFT10M              &  10M     &  10K     &  10M       &  128   & Euclide         &       Image                   \\ \hline
DEEP10M              &  10M     &  10K     &  10M       &  96   & Cosine         &     Image                     \\ \hline
\end{tabular}
\caption{Statistics of the datasets. $|X|$, $|Q|$ and $|T|$ denote the number of base data, test queries and historical queries, respectively. $d$ denotes the dimensionality of vectors.}
\label{dataset}
\vspace{-2em}
\end{table}
\subsection{Experimental Setup}
\textbf{Datasets.} We mainly evaluate our method using four modern large-scale cross-modal datasets. In Text-to-Image~\cite{t2i} and LAION~\cite{laion}, the base data are image vectors, while the queries are text vectors. In WebVid~\cite{webvid}, the base data are embeddings of videos, and the queries are text vectors. We applied deduplication to the queries, ensuring that the test queries are entirely different from historical queries. We then randomly sampled 10k queries that are different from historical queries for testing. Specifically, the MainSearch dataset is sourced from a large e-commerce platform, where the base data corresponds to products. Each product comprises an image and accompanying text, which we combine and embed into a single vector. Query vectors, on the other hand, are derived from either text or image embeddings. The queries are collected over a continuous time, and we use the last 50k queries that differ from historical queries for testing. For the MainSearch dataset, we primarily evaluate the performance of different graph indexes when historical queries are limited and the performance of our method when constructing the index online. Additionally, we conduct experiments using two single-modal datasets~\cite{pq, deep__}. The statistics of the datasets are listed in Table \ref{dataset}. 

In the Text-to-Image dataset, text and image embeddings are generated by a variant of DSSM~\cite{dssm} and SE-ResNeXt-101~\cite{se_resnext}, respectively, and are mapped into a shared space via contrastive learning. The embeddings in LAION and WebVid are generated using CLIP-ViT-B/32~\cite{clip}. In MainSearch, we first use a 3B-parameter multimodal LLM \footnote{Due to commercial privacy considerations, we are unable to provide further details about the LLM.} to generate 3072-dimensional embeddings, which are then compressed to 256 dimensions using Matryoshka Representation Learning~\cite{mrl}. The embeddings in DEEP and SIFT datasets are generated by GoogLeNet~\cite{googlenet} and Scale-Invariant Feature Transform~\cite{sift_model}, respectively.

\textbf{Algorithms and Parameter Setting.} We use $\mathbb{G}$-NGFix to denote the graph index after NGFix on base graph $\mathbb{G}$ and we use $\mathbb{G}$-NGFix* to denote the graph index after NGFix and RFix on $\mathbb{G}$. For cross-modal datasets, since RoarGraph has significantly outperformed existing graph algorithms~\cite{roar_graph}, we mainly compare the performance of HNSW-NGFix* and RoarGraph~\cite{roar_graph}, using HNSW~\cite{hnsw} and NSG~\cite{nsg} as baselines. For single-modal datasets, we compare the performance of HNSW-NGFix*, HNSW, $\tau$-MNG~\cite{t_mg}, RoarGraph and NSG. We determine the parameters for each graph index through multiple experiments. The parameters are as follows: (1) \uline{HNSW-NGFix*}: We set $M_{EX}=48$ to limit the out-degree of each node. For each historical query, we apply NGFix and RFix twice to fix defects in the graph. In the first round of fixing, we set $N_q=100$, $K_h=100$ and $MaxS \le 500$; in the second round, we set $N_q=10$, $K_h=10$ and $MaxS \le 50$. The reason for performing two rounds of fixing is that using $N_q=100$ mainly improves performance when the search list size $L$ is large (i.e., $\ge 100$). However, in scenarios where only a small number of nearest neighbors are required (e.g., retrieving the top 10 NNs), a smaller $L$ is sufficient. To ensure performance in such cases, we apply NGFix and RFix again with $N_q=10$ \footnote{For RFix, setting $N_q=10$ also ensures that the search can reach the vicinity of the query when $N_q=100$. Therefore, RFix only needs to be performed once with $N_q=10$.}. The number of historical queries used for NGFix and RFix is summarized in Table \ref{dataset}. We use brute-force to calculate the exact $k$NN of each historical query. For cross-modal datasets, to prevent excessive node out-degrees, we set $M=16$ and use $efC=2000$ to obtain high-quality neighbors for each point in base graph. In contrast, for single-modal datasets, where the number of hard queries is relatively small and NGFix adds fewer edges, we set $M=32$ and $efC=2000$. Existing research indicates that in uniformly distributed random data, the upper layers of HNSW play a limited role when the dimensionality $d \geq 32$~\cite{hnsw_down_h}. In vector retrieval scenarios, $d$ typically exceeds $32$, and the upper layers of HNSW also demonstrate limited effectiveness under real-world data distributions~\cite{hnsw_h_useless,hnsw_h_useless2,steiner_hardness}. To avoid the additional indexing time and size overhead caused by the upper layers, we use only the bottom layer of HNSW as base graph $\mathbb{G}$.
(2) \uline{RoarGraph}: Similar to $\mathbb{G}$-NGFix, we set $N_q=100$ and use the same number of historical queries as $\mathbb{G}$-NGFIX. Additionally, we set $M=32$ and $L=2000$ to obtain high-quality neighbors. 
(3) \uline{HNSW}: We set $efC=2000$ to provide good neighbors for nodes in the graph, and we set $M=32$ to control the out-degree of each node. 
(4) \uline{NSG}: We set $C=2000$ and $L=2000$ to provide good quality of neighbors, and we set $R=64$ to control the average out-degree of the graph. 
(5) \uline{$\tau$-MNG}: For the same reasons as NSG, we set $R=64$, $C=2000$, $L=2000$. According to the recommendations in the paper, we set $\tau$ to $10$ for SIFT10M and $\tau$ to $0.01$ for DEEP10M.

\textbf{Evaluation metrics.}
Following previous works~\cite{roar_graph, ood_diskann, t_mg, nsg, hnsw}, we use average recall@$k$ and rderr@$k$ (defined in Section \ref{preliminaries}) to measure the query accuracy. We use query-per-second (QPS) and Number of Distance Calculations (NDC) to measure efficiency. We compare the QPS at the same recall@$k$ and the NDC at the same rderr@$k$, where a higher QPS or lower NDC indicates a better index. When plotting the recall@$k$-QPS curves, we initially set search list size $L$ (a.k.a., ef\_search in HNSW) to $k$, and then incremented it by 10 at each step. For each $L$, we recorded the corresponding QPS and recall@$k$. We select representative points to plot the curves. The rderr@$k$-NDC curves were generated in the same manner. We mainly focus on the cases where $k=10$ and $k=100$.

We implemented NGFix* in C++ and compiled our code using g++ 10.1.0. Experiments were conducted on an Intel(R) Xeon(R) Gold 5220 CPU@2.20GHz with 256GB DDR4 memory (@2666MT/s) under CentOS7.9. We used the same distance computation functions for different graph indexes. All queries were performed with a single thread. Additionally, we utilize 32 threads when evaluating the index construction time of different indexes.

\subsection{Search performance}

\begin{figure*}\centering
        \setlength{\abovecaptionskip}{0.1cm}
	\includegraphics[width=5.4in]{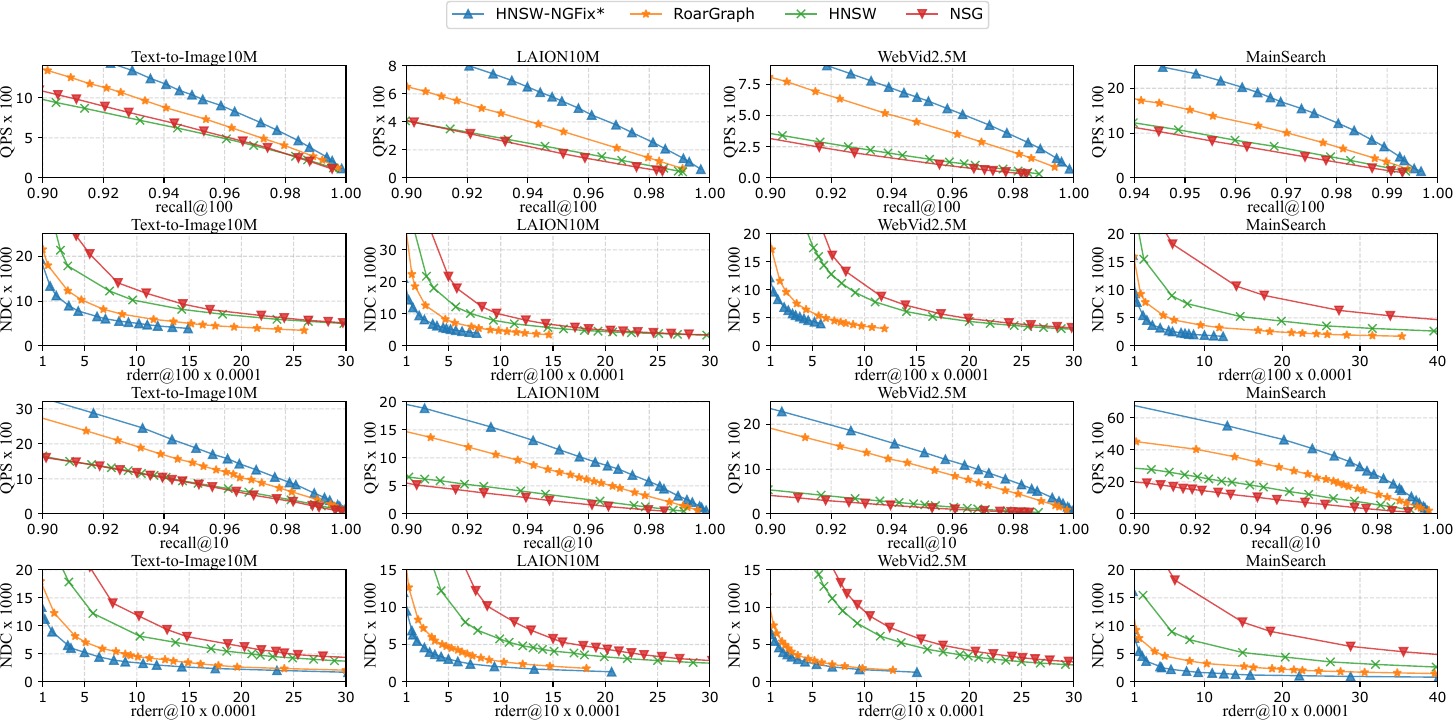}
	\caption{Search performance on cross-modal datasets.}
    \Description{..}
    \label{cross-modal_result}
    % \vspace{-1em}
\end{figure*}

\textbf{Cross-Modal Datasets.} Figure \ref{cross-modal_result} presents the QPS–recall@100 and NDC–rderr@100 curves for various graph indexes across four cross-modal datasets. HNSW-NGFix* outperforms the other algorithms: at recall@100=0.95, its QPS is 1.3–1.6 times that of RoarGraph and 1.7–3.66 times that of HNSW. At recall@100=0.99, HNSW-NGFix*’s advantage becomes even more pronounced, with a QPS 1.31–2.25 times higher than RoarGraph and 1.78–6.88 times higher than HNSW. In terms of NDC and rderr@100, HNSW-NGFix* also demonstrates significant superiority, especially when aiming for low error rates. When rderr@100 is less than 0.0001, HNSW-NGFix* requires only 48.5\%–78\% of the distance calculations needed by RoarGraph. For the case of $k=10$, HNSW-NGFix* also achieves better search performance compared to other graph indexes. Besides the performance improvement, we have the following additional insights. (1) On the MainSearch dataset, we can still achieve significantly higher performance than HNSW using only a small number of historical queries. This is because queries in real-world production environments often exhibit similar characteristics, making it more likely for the test queries to fall within regions of the graph where defects have already been fixed. (2) Compared to RoarGraph, HNSW-NGFix* shows more significant performance advantages when achieving high recall (e.g., recall@100 > 0.98) or low rderr (e.g., rderr@100 < 0.0001). The reason lies in our use of EH to guide graph construction, which takes advantage of the hard queries with poor neighboring structures. The edges linked by these queries make the final graph index more effective in achieving high recall.

\begin{figure}
    \centering
    \setlength{\abovecaptionskip}{0.1cm}
    \includegraphics[width=0.7\linewidth]{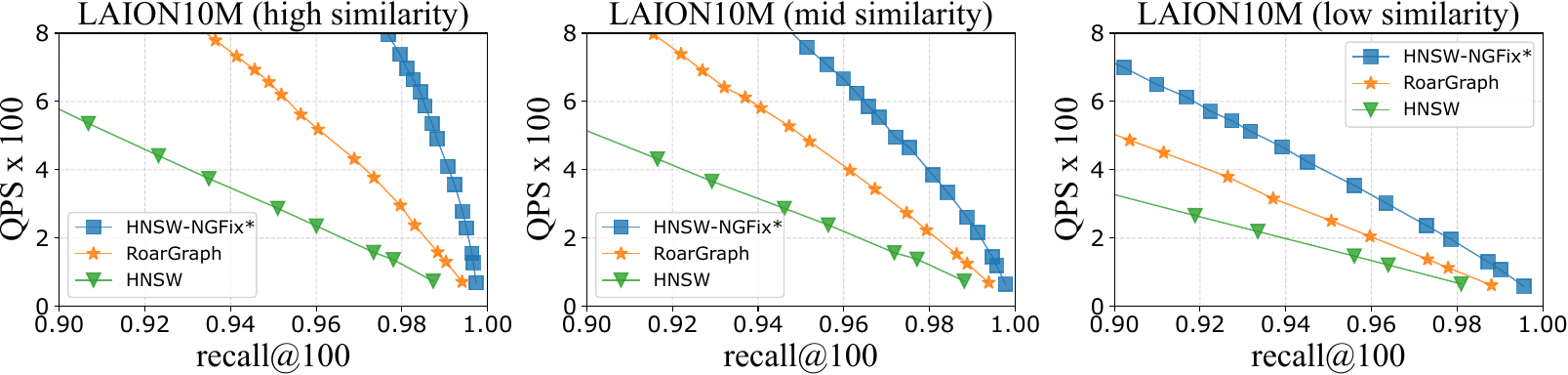}
    \caption{Evaluation of queries with different similarities.}
    \Description{..}
    \label{similarities}
    % \vspace{-1em}
\end{figure}

\textbf{OOD Queries with Different Similarities.} We evaluate our method on test queries with varying similarity to historical queries. We measure query similarity using the distance $d$ between test queries and their nearest historical queries. In LAION10M (distance is defined as 1-cosine), 18.1\% of the test queries have $d \le 0.05$ (i.e., high similarity), and 26.2\% satisfy $0.05 < d \le 0.1$ (i.e., moderate similarity). The remaining queries exhibit low similarity. The thresholds depend on the characteristics of each dataset. For LAION10M, we choose 0.05 and 0.1 to better illustrate the relationship between index performance and query similarity. Figure \ref{similarities} demonstrates that our method outperforms others under varying similarity levels.

\begin{figure}
    \centering
    \setlength{\abovecaptionskip}{0.1cm}
    \includegraphics[width=0.7\linewidth]{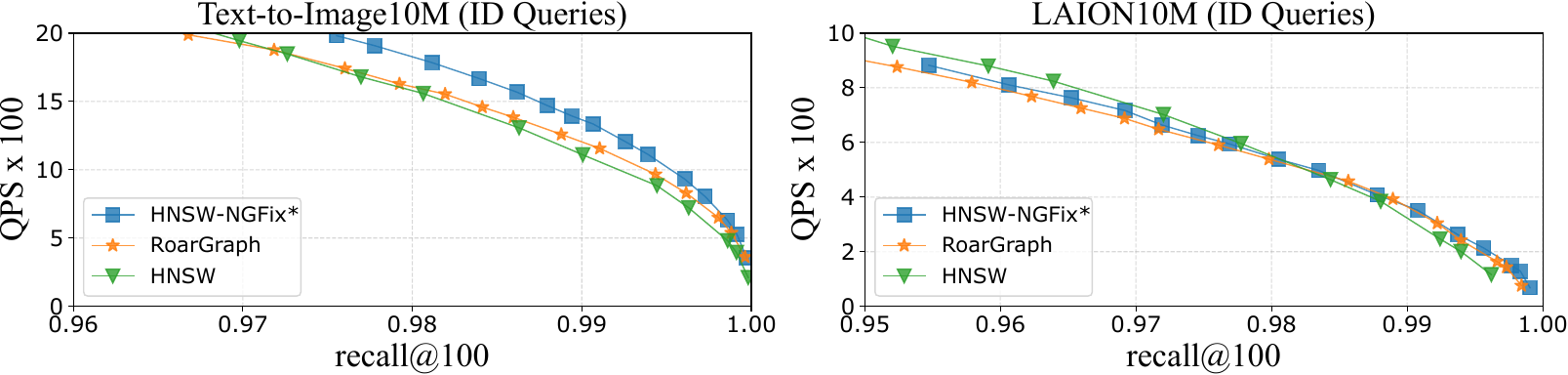}
    \caption{Evaluation of ID Queries in cross-modal datasets.}
    \Description{..}
    \label{id_query_cross}
    % \vspace{-1em}
\end{figure}

\textbf{ID Queries in Cross-Modal Datasets.} In cross-modal retrieval, both OOD queries (e.g., text-to-image) and ID queries (e.g., image-to-image) may occur. As shown in Figure \ref{id_query_cross}, the index refined by NGFix* with OOD queries also achieves strong performance on ID queries. This indicates that fixing with OOD queries does not affect the performance of ID queries.

\begin{figure}
    \centering
    \setlength{\abovecaptionskip}{0.1cm}
    \includegraphics[width=0.7\linewidth]{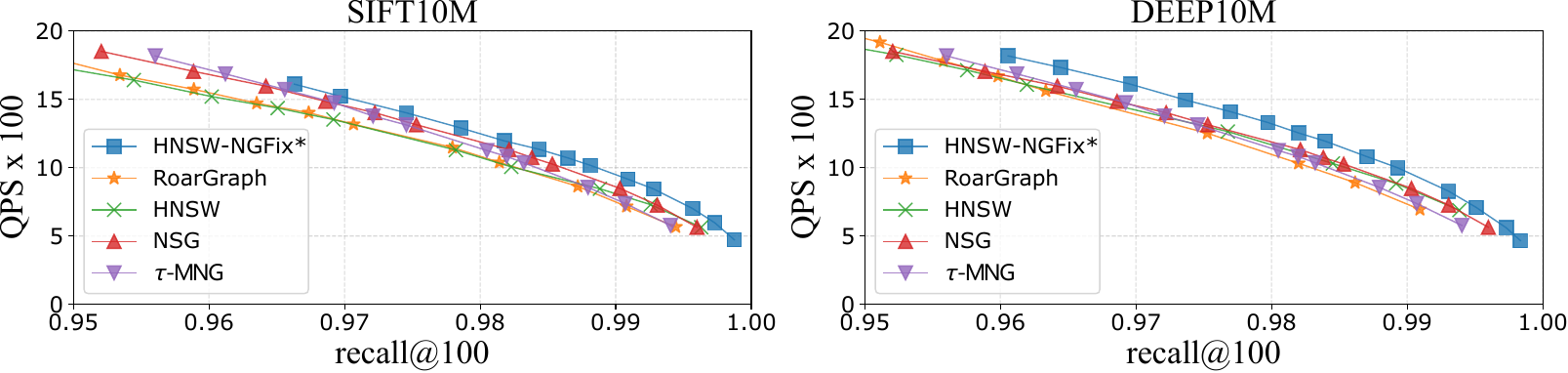}
    \caption{Search performance on single-modal datasets.}
    \Description{..}
    \label{single-modal_result}
    % \vspace{-1em}
\end{figure}

\textbf{Single-Modal Datasets.} Figure \ref{single-modal_result} shows the QPS-recall@100 curves in single-modal datasets. Our approach achieves modest QPS improvements of approximately 10\%. This limited improvement is due to NGFix's emphasis on hard queries, which are rare in single-modal datasets, leading to fewer effective edge connections. Another reason is that HNSW and NGFix* construct the graph based on the base data distribution and the query distribution, respectively. However, for single-modal datasets, these two distributions are consistent. Since HNSW already fits the base data distribution well and only a few hard queries that are relatively far from the base data benefit from our method, the improvement brought by NGFix* is limited in this case. On the other hand, RoarGraph's graph construction strategy fails to leverage the information provided by these hard queries. In some cases, the edges even increase the computational burden during the search, causing its performance to be even worse than HNSW.

Since the single-modal datasets contain fewer hard queries and we can easily achieve high accuracy, our subsequent experiments primarily focus on evaluating the performance of our method on cross-modal datasets.

\subsection{Effect of Historical Query Set Size.}
\begin{figure}
    \centering
    \setlength{\abovecaptionskip}{0.1cm}
    \includegraphics[width=0.7\linewidth]{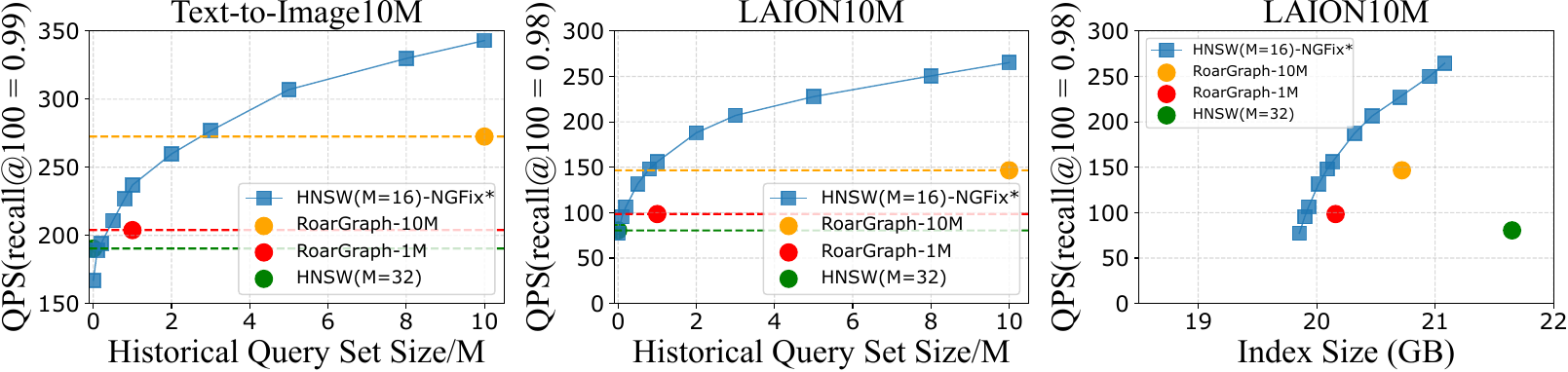}
    \caption{Evaluation of historical query set size.}
    \Description{..}
    \label{train_size}
    % \vspace{-1em}
\end{figure}

In this experiment, we evaluate the performance of our method with varying numbers of historical queries. Since RoarGraph needs to entirely rebuild the graph when the number of historical queries differs from the previous configuration, building RoarGraph with different quantities of historical queries demands substantial time. So we only selected RoarGraph with 1M and 10M historical queries for comparison and we denote RoarGraph with $p$ historical queries as RoarGraph-$p$. Figure \ref{train_size} illustrates the QPS achieved by HNSW-NGFix* at specific recall@100 across different numbers of historical queries. HNSW-NGFix* requires only 8\%-30\% of the historical queries compared to RoarGraph-10M when achieving the same performance on different datasets. Additionally, HNSW (M=16)-NGFix*, using merely 1\% of the historical queries relative to the base data size, achieves the same performance as HNSW (M=32). These results demonstrate that our method can achieve superior performance with fewer historical queries and can attain high performance in a short time when constructing the graph online. The rightmost subfigure in Figure \ref{train_size} shows the trade-off between index size and QPS. We vary the number of historical queries used to control the index size of HNSW-NGFix*. The results indicate that HNSW-NGFix* achieves higher QPS compared to both RoarGraph and HNSW under the same index size.

\subsection{Ablation Study}\label{sec:appro_gt}
\begin{figure}
    \centering
    \setlength{\abovecaptionskip}{0.1cm}
	\begin{subfigure}{0.2\linewidth}
		\centering
		\includegraphics[width=1\linewidth]{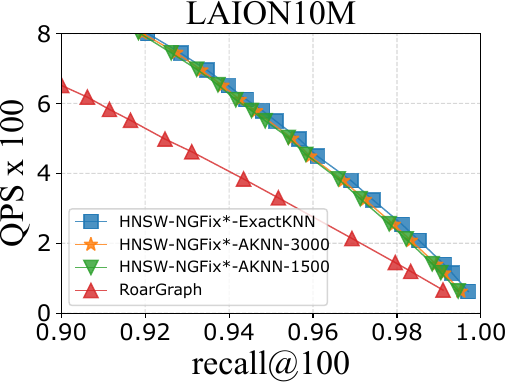}
        % \vspace{-1.5em}
		\caption{}
		\label{preprocess}
	\end{subfigure}
	\begin{subfigure}{0.225\linewidth}
		\centering
		\includegraphics[width=1\linewidth]{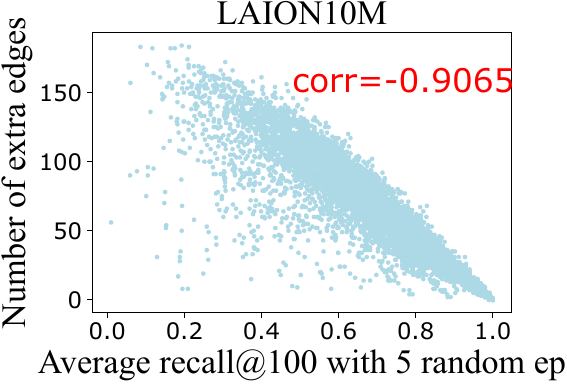}
        % \vspace{-1.5em}
		\caption{}
		\label{eh_corr}
	\end{subfigure}
    \begin{subfigure}{0.2\linewidth}
		\centering
		\includegraphics[width=1\linewidth]{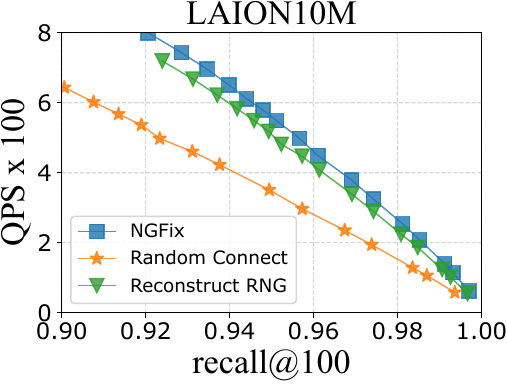}
        % \vspace{-1.5em}
		\caption{}
		\label{fix}
	\end{subfigure}
    \caption{(a) Evaluation of different preprocessing methods. (b) Evaluation of the Escape Hardness. (c) Evaluation of different defects fixing methods.}
    \Description{...}
    \label{ablation1}
    % \vspace{-1em}
\end{figure}

\textbf{Evaluating the impact of different preprocessing methods.} We compare the graph quality achieved by using exact $k$NN against that of approximate $k$NN (described in section \ref{sec:preprocess}). HNSW-NGFix-ExactKNN represents the graph constructed using exact $k$NN obtained through brute-force methods for historical queries. Conversely, HNSW-NGFix-AKNN-$L$ denotes the graph built using approximate $k$NN derived from a Greedy Search with a search list size of $L$. As illustrated in Figure \ref{preprocess}, our experimental results show that the QPS-recall@100 curves for approximate $k$NN are almost identical to those of exact $k$NN, with only a slight decrease.

\textbf{Evaluating the effectiveness of EH in differentiating between easy and hard queries.} We evaluate whether Escape Hardness could guide us in linking different numbers of edges according to the hardness of the queries. Figure \ref{eh_corr} reveals a high correlation between query accuracy and the number of additional edges connected by NGFix. This indicates that our algorithm concentrates on hard queries by adding more edges for them while linking fewer edges for easy queries. Consequently, our approach not only optimizes the graph index size but also reduces the overhead of distance calculations during searches.

\textbf{Evaluating different defects fixing methods.} Figure \ref{fix} illustrates the performance of graphs generated by different defects fixing methods (introduced in Figure \ref{defects_fix_different_example}). Among these, random connecting exhibits the poorest performance because randomly connected edges fail to provide effective neighbors for each node. Meanwhile, NGFix achieves better performance than the method of reconstructing RNG. Moreover, the average out-degree of the graph derived from Reconstructing RNG is approximately 1.37 times that of NGFix, indicating that NGFix reduces the index size while maintaining a high-quality graph structure.

\begin{figure}
    \setlength{\abovecaptionskip}{0.1cm}
    \centering
    \includegraphics[width=0.7\linewidth]{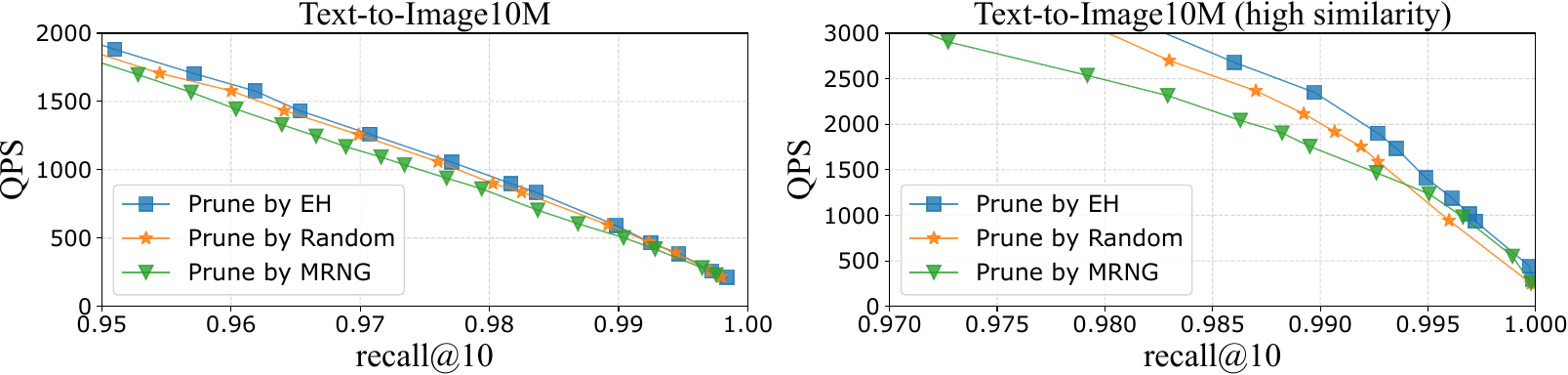}
    \caption{Evaluation of different edge-pruning strategies.}
    \Description{..}
    \label{prune}
    % \vspace{-1em}
\end{figure}

\textbf{Evaluating the impact of different edge-pruning strategies. }
Figure \ref{prune} illustrates the performance of different edge-pruning strategies employed in NGFix (Algorithm \ref{defects_fix} lines 11-15). Pruning based on EH achieves better performance than random pruning, whereas MRNG pruning exhibits the poorest performance. This is because MRNG pruning is designed for queries that overlap with the base data~\cite{nsg} and is more likely to prune long edges (i.e., edge $(u, v)$ with large $\delta(u,v)$). However, since the $k$NN of hard queries are often scattered across different regions~\cite{roar_graph}, these long edges are often very useful for hard queries, and pruning them adversely affects the connectivity of the graph around hard queries, leading to significant losses in accuracy and performance.

\begin{figure}
    \setlength{\abovecaptionskip}{0.1cm}
    \centering
    \includegraphics[width=0.7\linewidth]{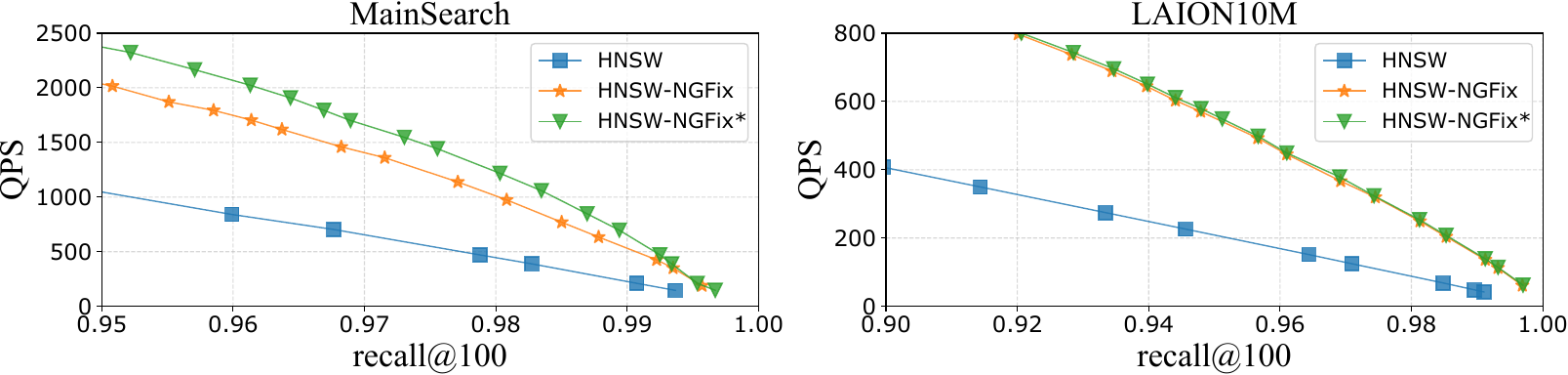}
    \caption{Evaluation of Reachability Fixing}
    \Description{..}
    \label{rf_test}
    % \vspace{-1em}
\end{figure}

\textbf{Evaluating the impact of the RFix.} Figure \ref{rf_test} shows the performance of NGFix and NGFix*. In the MainSearch dataset, due to the relatively large proportion of searches that cannot reach the query vicinity (Figure \ref{recall}), NGFix* improves performance by 18\% over NGFix when recall@100=0.95. In LAION10M, since the search for almost all queries can reach the query vicinity before RFix, NGFix* only shows a slight improvement over NGFix.

\subsection{Construction Time and Index Size}

\begin{figure}
    \setlength{\abovecaptionskip}{0.1cm}
    \centering
    \includegraphics[width=0.7\linewidth]{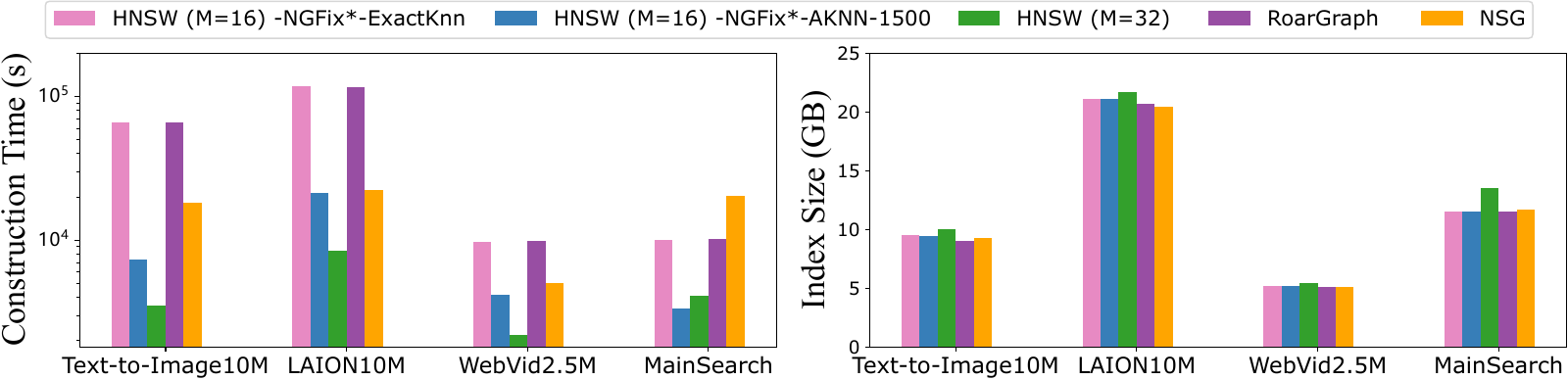}
    \caption{Construction time and index size}
    \Description{..}
    \label{index_size}
    % \vspace{-1em}
\end{figure}

Figure \ref{index_size} reports the index construction time and the index size across different datasets. The index size reflects the memory consumption during the search. When using the exact $k$NN for graph construction, the indexing time of HNSW-NGFix* is almost the same as that of RoarGraph. However, when a large number of historical queries are involved, the construction time becomes significantly higher than that of other methods. When using approximate $k$NN, HNSW-NGFix* accelerates index construction by 2.35-9.02 times compared to RoarGraph (RoarGraph lacks a complete graph index when using historical queries, which makes it unable to leverage approximate $k$NN for graph construction). Meanwhile, HNSW-NGFix* requires 82.1\% to 210.2\% of the indexing time taken by HNSW, and 16.5\% to 95.5\% compared to NSG. On the MainSearch dataset, HNSW achieves its best performance with $M=32$, while HNSW-NGFix* only needs $M=16$, resulting in a shorter index construction time than HNSW. In terms of index size, HNSW-NGFix* is smaller than HNSW since it only utilizes the bottom layer of HNSW. However, as we store the EH of each edge (with an additional 16 bits per extra edge) for edge pruning, the index size of HNSW-NGFix* is slightly larger than that of RoarGraph and NSG.

\subsection{Effect of Different Parameters}
\label{sec::para}
\begin{figure}
    \setlength{\abovecaptionskip}{0.1cm}
    \centering
    \includegraphics[width=0.7\linewidth]{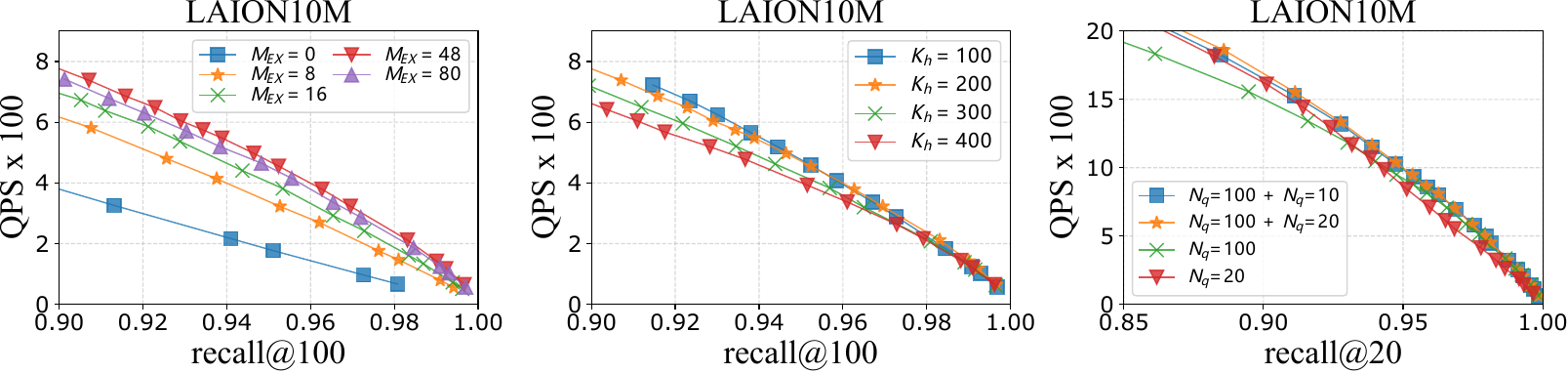}
    \caption{Evaluation of different parameters.}
    \Description{..}
    \label{para_test}
    % \vspace{-1em}
\end{figure}
Figure \ref{para_test} illustrates the performance of the graph index under different parameter settings. \uline{For the parameter $M_{EX}$, } we fix $K_h=200$ and perform a single NGFix with $N_q=100$ to evaluate the impact of $M_{EX}$. Since a smaller $M_{EX}$ leads to the pruning of useful edges, while a larger $M_{EX}$ increases the search overhead, we recommend setting $M_{EX}$ in the range of 32–64. For datasets with higher average query hardness (as measured by EH), a larger $M_{EX}$ may be appropriate. Nevertheless, setting $M_{EX}=48$ generally yields sufficiently good performance. \uline{For the parameter $K_h$}, we fix $M_{EX}=48$ and also perform a single NGFix with $N_q=100$ to evaluate the impact of $K_h$. Experimental results show that a smaller $K_h$ tends to produce a more effective index, but at the cost of increased index size. For example, when $K_h=100$ and $K_h=400$, the average out-degrees are 41.26 and 30.85, respectively. Therefore, $K_h$ can be selected based on the trade-off between space and performance requirements. \uline{For the parameter $N_q$}, we recommend first setting $N_q=100$, and then performing NGFix* again with $N_q=k$ based on the number of NNs $k$ to be retrieved. The experimental results in the rightmost subfigure of Figure \ref{para_test} show that when $k=20$, performing NGFix* with $N_q=20$ yields slightly better performance. Nonetheless, in scenarios where $k$ varies frequently, it is not necessary to perform NGFix* for all values of $k$. Performing NGFix* twice with $N_q=10$ and $N_q=100$ is sufficient to produce a high-quality index. We include more details about $N_q$ in Appendix C. \uline{For the parameter $MaxS$}, since very few edges have an EH value exceeding $5N_q$, setting $MaxS$ to $5N_q$ is adequate. If $N_q$ is set to a large value (e.g., $\ge 300$), it is recommended to set $MaxS$ to 1.2$N_q$–2$N_q$ to reduce the index construction overhead, as lowering $MaxS$ will not significantly affect the index quality. \uline{For the parameter settings of the base graph}, we can follow the selection strategy of the corresponding graph index. Meanwhile, slightly reducing the average out-degree is recommended to avoid an overly large final graph (e.g., reducing HNSW's $M$ by 16).

\subsection{Evaluation of Index Maintenance Methods}\label{sec:index_maintenance}
\begin{figure}
    \setlength{\abovecaptionskip}{0.1cm}
    \centering
    \includegraphics[width=0.7\linewidth]{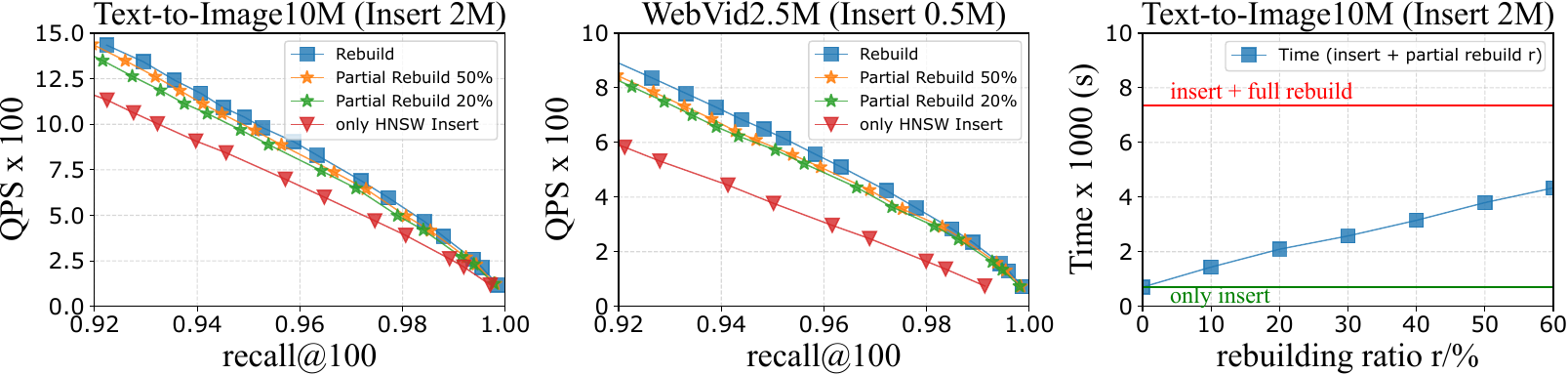}
    \caption{Evaluation of insertion method}
    \Description{..}
    \label{insert}
    % \vspace{-1em}
\end{figure}

\begin{figure}
    \setlength{\abovecaptionskip}{0.1cm}
    \centering
    \includegraphics[width=0.7\linewidth]{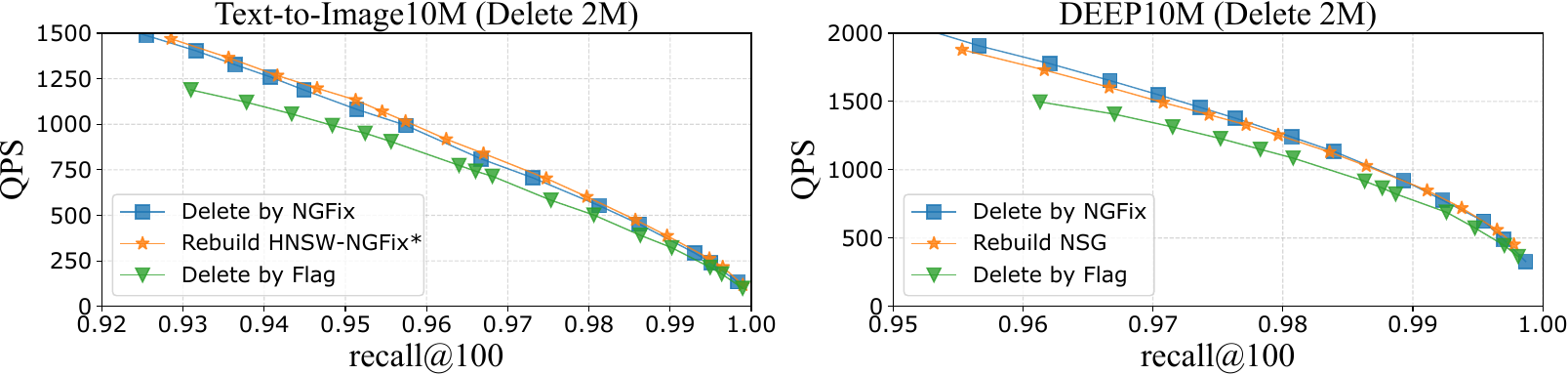}
    \caption{Evaluation of deletion method}
    \Description{..}
    \label{delete}
    % \vspace{-1em}
\end{figure}

In this experiment, to ensure a fair comparison with rebuilding, we perform both insertion and deletion operations using 32 threads.

\textbf{Insertion.} Figure \ref{insert} shows the performance of the graph index after inserting 20\% of the data points using different insertion algorithms. “Partial Rebuild $r$” in the figure denotes a partial rebuilding with proportion $r$, as described in Section \ref{sec:insertion}. We perform a partial rebuilding after inserting 20\% of the data. Compared with directly using HNSW's insertion algorithm, partial rebuilding significantly improves the quality of the resulting graph index. A higher proportion leads to better index quality. The rightmost subfigure in Figure \ref{insert} shows the relationship between time (insertion + rebuilding) and the parameter $r$. When $r=20\%$, the total time is only 28.5\% of that for full rebuilding. In practice, we can choose an appropriate proportion based on time and quality requirements. If new points are inserted during partial rebuilding, the insertion and rebuilding processes remain independent. This is because partial rebuilding only modifies the extra edges, while inserting new points only modifies the base graph. So the total insertion time is not affected.

\textbf{Deletion.} When performing deletions with NGFix, we adopt greedy search with a search list size of 800 to obtain the approximate ground truth, and set parameters $N_q=100$, $K_h=100$, $MaxS \le 500$. Figure \ref{delete} presents the results of the deletion experiment. The left panel shows the performance of different methods after removing 20\% of the base data on the Text-to-Image dataset. It can be observed that the performance of the index degrades significantly under the lazy deletion strategy. In contrast, the index quality achieved by applying NGFix after deletion is nearly identical to that of full reconstruction, with only a slight drop. The time cost of deletion using NGFix is only 6.8\% of that of a full rebuilding. To demonstrate the robustness of our deletion algorithm, the right panel shows the performance after deletion using NGFix on an NSG index (without NGFix* based on historical queries). It can be observed that the performance after deletion even surpasses that of full reconstruction, as NGFix is capable of linking edges of higher quality than those in the original NSG. This demonstrates that our proposed deletion method also performs well on other graph indexes.

\section{DISCUSSION AND FUTURE WORK}

\begin{figure}
    \setlength{\abovecaptionskip}{0.1cm}
    \centering
    \includegraphics[width=0.7\linewidth]{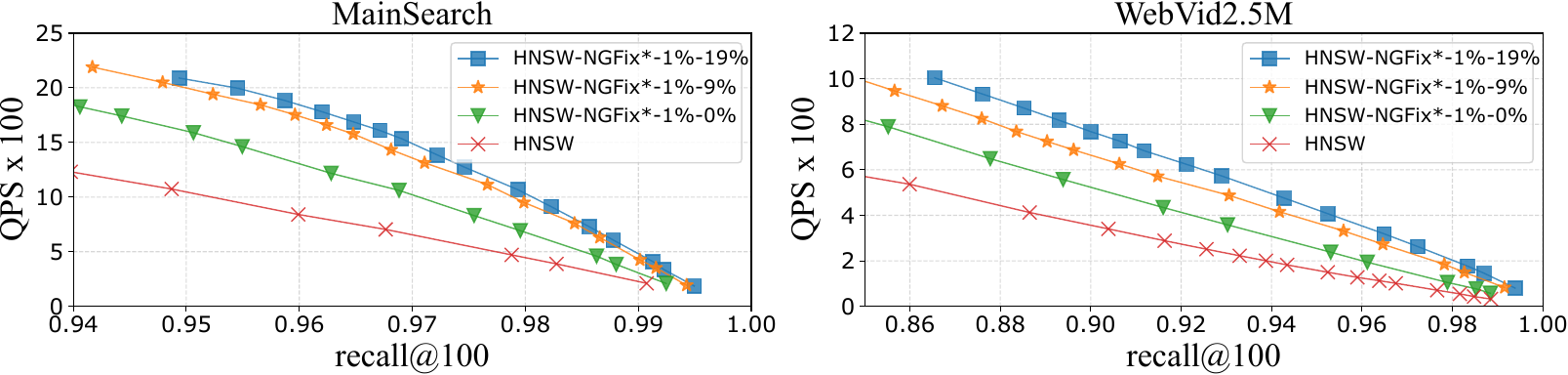}
    \caption{Evaluation of generated historical queries.}
    \Description{..}
    \label{cold_start}
    % \vspace{-1em}
\end{figure}

\textbf{Mitigate the dependence on historical queries. }In certain scenarios (e.g., cold-start), the lack of sufficient historical queries may degrade index performance. We propose two strategies to mitigate this issue: (1) Since multimodal embedding models typically require a large amount of multimodal data for training, we can use the training data as historical queries. (2) In scenarios involving workload drift, only a few representative queries may be available. To address this limitation, we propose a data augmentation method to synthetically generate a larger set of queries. We define HNSW-NGFix*-p\%-q\% to use some real historical queries equal to p\% of the base data size, along with generated queries equal to q\% of the base data size. For each real historical query, we generate q/p synthetic queries. In each generation, Gaussian noise with zero mean and variance $\sqrt{c/d}$ is added independently to every dimension of historical queries ($d$ represents the dimensionality). In experiments on WebVid and MainSearch, we selected $c$ = 0.3 as it achieved the best performance among $c$ = 0.1, 0.2, 0.3, 0.4. Experimental results in Figure~\ref{cold_start} demonstrate the effectiveness of this method.

\textbf{Selection strategy for historical queries. }When the query workload remains stable, it is unnecessary to design a historical query selection strategy, as NGFix* can adaptively adjust the graph index based on the query hardness. For queries with low EH, NGFix* adds very few edges or skips them altogether. In contrast, for high-EH queries, it adds more edges to effectively fix defects in the graph. So it is sufficient to use all available historical queries. In scenarios with dynamic workloads, the limited out-degree of each node may prevent new edges from being added for incoming queries, as edges added for old historical queries may occupy the out-degree budget. To prioritize fixing the graph structure for new queries, we periodically delete a subset of existing extra edges added by NGFix* to make room for newly added edges (e.g., by randomly selecting 20\% extra edges). Then, we prioritize using the newest queries based on their timestamps to fix the graph. This allows the index to better adapt to the new workload. Exploring better strategies to adapt to workload changes could be considered a direction for future work.

\begin{figure}
    \centering
    \setlength{\abovecaptionskip}{0.1cm}
    \begin{subfigure}{0.21\linewidth}
		\centering
		\includegraphics[width=0.7\linewidth]{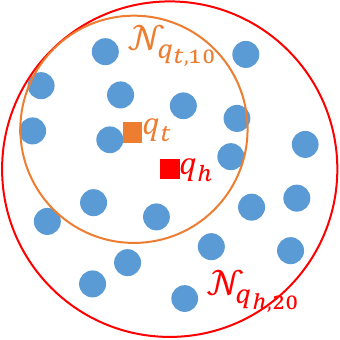}
		\caption{}
		\label{theory_example}
	\end{subfigure}\hspace{3mm}
	\begin{subfigure}{0.64\linewidth}
		\centering
		\includegraphics[width=0.7\linewidth]{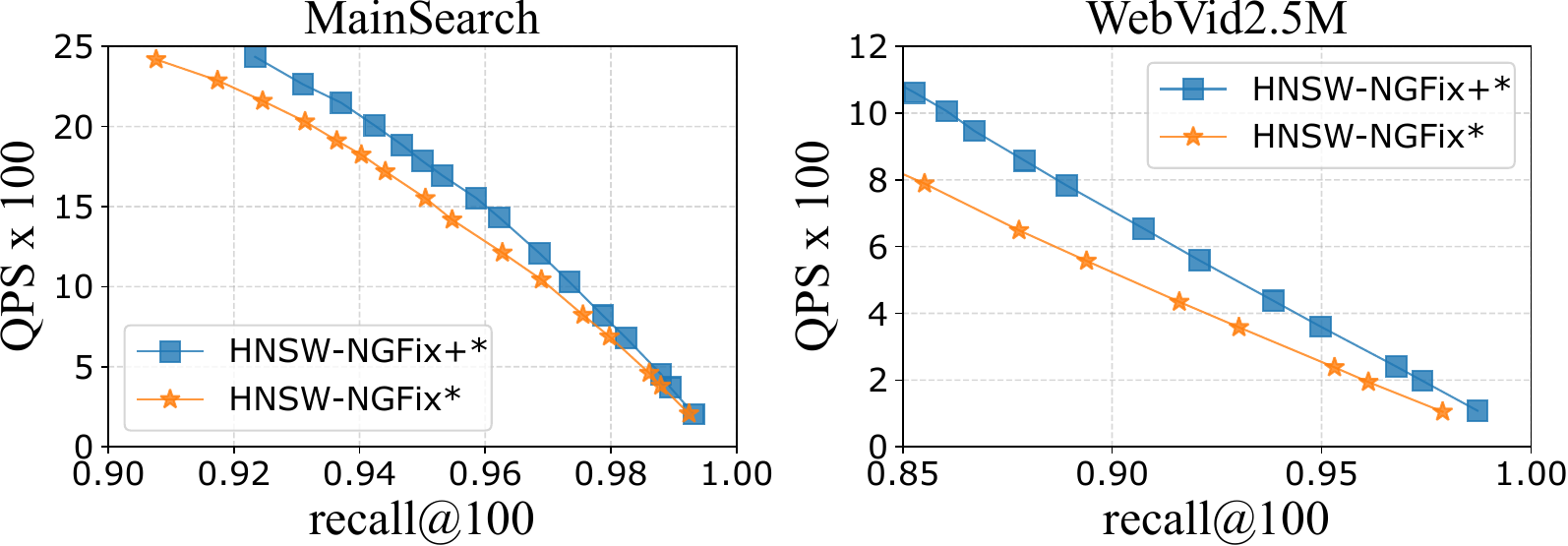}
		\caption{}
		\label{ngfix+}
	\end{subfigure}
    \caption{(a) Example of NGFix+. (b) Evaluation of NGFix+.} 
    \Description{...}
    \label{fig3}
    % \vspace{-1em}
\end{figure}

\textbf{Extending the theoretical guarantee. } Let $\mathcal{N}_{q,k}$ denote the set of the top-$k$ NN of query $q$. Let $q_h$ denote historical query, and $q_t$ denote the new test query. One direction for extending the theoretical guarantee is to ensure query accuracy when $\delta(q_h, q_t) < \varepsilon$. We let $U_{k}$ denote the union of the top-$k$ NN sets of all queries $q_t$ that satisfy $\delta(q_h, q_t) < \varepsilon$. For each $q_h$ and a fixed $\epsilon$, in order to guarantee the accuracy of all queries satisfying $\delta(q_h, q_t) < \epsilon$, we need to consider all points in the set $U_k$. Since $|U_k|$ can be very large (in the worst case, it could be the size of the base data), ensuring query accuracy becomes very difficult. Therefore, we try to extend the theoretical guarantee from another perspective. Assuming that the search can reach the vicinity of the query, one direction for theoretical extension is as follows: \textbf{when $\mathcal{N}_{q_t,k} \subseteq \mathcal{N}_{q_h, K}$, we aim to guarantee that the recall@$k$ for $q_t$ is 100\%, where search list size $L\ge k$, $K = ck$ and $c$ is a constant (e.g., 2)}. Figure \ref{theory_example} provides an example of this theory. Our method (NGFix) is already able to ensure accuracy where $c=1$ (i.e., $\mathcal{N}_{q_t,k} = \mathcal{N}_{q_h,k}$). In the WebVid dataset, experimental results show that when $\mathcal{N}_{q_t,10} = \mathcal{N}_{q_h,10}$, $\delta(q_t, q_h)$ is approximately less than 0.03 (Euclidean distance). Moreover, when $\mathcal{N}_{q_t,10} \subseteq \mathcal{N}_{q_h,20}$, $\delta(q_t, q_h)$ is approximately less than 0.114. Details of this experiment are provided in Appendix D.

To ensure the property when $c>1$, a trivial approach is to enumerate all $\mathcal{N}_{q_t,k}$ and apply NGFix for each one. We conducted experiments on cross-modal datasets with parameters set to $k=100$ and $K=200$ (i.e., $c=2$). To avoid excessive graph construction overhead, we use a small number of historical queries ($2.5$$\times$$10^4$ in WebVid and $10^5$ in MainSearch) and randomly enumerate 100$\mathcal{N}_{q_t,100}$ for each historical query. We denote the new method as NGFix+, and the experimental results in Figure \ref{ngfix+} show that NGFix+ outperforms NGFix. However, NGFix+ takes 19.2 times longer to run than NGFix. Therefore, one direction for future work is to develop more efficient algorithms to ensure the above property.

\textbf{Hash-Table Method. }When test queries exactly overlap with historical queries, we can use a hash-table method. The main steps are as follows: (1) Apply a hash function (e.g., MD5~\cite{md5}) to each query to obtain a compact key, and store the key along with the corresponding ground truth for historical queries in a hash table. (2) For a test query, if it can be retrieved from the hash table, directly return the ground truth; otherwise, use ANNS. The latency of the hash-table method is about 9.3\% of that of graph-based ANNS in MainSearch dataset (using HNSW-NGFix* with search list size $L=100$). However, the hash-table method cannot be generalized to unseen queries. Moreover, this method requires substantial extra space to store the ground truth and may suffer from errors due to hash collisions. In scenarios with high query repetition, using both hash table and HNSW-NGFix* can achieve higher QPS.

\textbf{Query Similarities. }In graph-based ANNS, a fixed search parameter $L$ is typically used for all queries when aiming for a predetermined target recall~\cite{hnsw,nsg,roar_graph,symphonyQG}. However, the experimental results in Figure \ref{similarities} show that the value of $L$ required (i.e., the latency required) to achieve a specific recall varies significantly across queries with different similarity levels. This observation suggests the potential for the following adaptive strategy: We first compute the similarity between the new query and historical queries, and then adjust the search parameter $L$ based on the similarity. Hence, one direction for future work is to develop methods for efficient query similarity calculation and fine-grained adjustment of $L$.

\section{CONCLUSION}
In this paper, we introduce NGFix and RFix, which are designed to improve graph index quality by exploiting the distribution of queries, with some theoretical foundations. Following prior work ~\cite{scc, steiner_hardness}, we divide the search process into two stages: (1) In the stage of approaching the query from the entry point, we analyze the limitations of existing approximation strategies for graph construction and introduce RFix to enhance the navigability of certain nodes; (2) In the stage of searching around the query, we first propose Escape Hardness to evaluate the quality of the graph index in the vicinity of the query. We then propose NGFix to dynamically fix defective regions of the graph based on Escape Hardness. Our method outperforms state-of-the-art methods on real-world datasets and significantly improves the accuracy of current hard queries.

\begin{acks}
We thank the anonymous reviewers for their valuable feedback and suggestions. This research is supported by the National Natural Science Foundation of China (grant numbers, 62272252, 62272253), and the Fundamental Research Funds for Central Universities.
\end{acks}

\bibliographystyle{ACM-Reference-Format}
\bibliography{sample-base}


\end{document}